\documentclass[aps,prb,twocolumn,amssymb,color,superscriptaddress,pdflatex]{revtex4-2}
\usepackage{color,physics}
\usepackage{epstopdf,xcolor,graphicx,hyperref,physics,amsmath,comment}

\date{\today}

\begin{document}
\title{Strain-induced superconductivity in Sr$_{2}$IrO$_{4}$}
\author{Lena Engstr\"om}
\email{lenae@physics.mcgill.ca}
\affiliation{Department of Physics and the Centre for the Physics of Materials, McGill University, Montr\'eal, Qu\'ebec,
	H3A 2T8, Canada}
\affiliation{D\'epartement de Physique, Universit\'e de Montr\'eal, Montr\'eal, Qu\'ebec, H3C 3J7, Canada}
\affiliation{Regroupement Qu\'eb\'ecois sur les Mat\'eriaux de Pointe (RQMP)}
\author{Chia-Chuan Liu}
\affiliation{D\'epartement de Physique, Universit\'e de Montr\'eal, Montr\'eal, Qu\'ebec, H3C 3J7, Canada}
\author{William Witczak-Krempa}
\affiliation{D\'epartement de Physique, Universit\'e de Montr\'eal, Montr\'eal, Qu\'ebec, H3C 3J7, Canada}
\affiliation{Regroupement Qu\'eb\'ecois sur les Mat\'eriaux de Pointe (RQMP)}
\affiliation{Centre de Recherches Math\'ematiques, Universit\'e de Montr\'eal, Montr\'eal, Qu\'ebec, HC3 3J7, Canada}
\affiliation{Institut Courtois, Universit\'e de Montr\'eal, Montr\'eal, Qu\'ebec, H2V 0B3, Canada}
\author{T.~Pereg-Barnea}
\affiliation{Department of Physics and the Centre for the Physics of Materials, McGill University, Montr\'eal, Qu\'ebec,
	H3A 2T8, Canada}
	\affiliation{Regroupement Qu\'eb\'ecois sur les Mat\'eriaux de Pointe (RQMP)}

\begin{abstract}
Multi-orbital quantum materials with strong interactions can host a variety of novel phases. In this work we study the possibility of interaction-driven superconductivity in the iridate compound Sr$_2$IrO$_4$ under strain and doping. We find numerous regimes of strain-induced superconductivity in which the pairing structure depends on model parameters. Spin-fluctuation mediated superconductivity is modeled by a Hubbard-Kanamori model with an effective particle-particle interaction, calculated via the random phase approximation. Magnetic orders are found using the Stoner criterion. The most likely superconducting order we find has $d$-wave pairing, predominantly in the total angular momentum, $J=1/2$ states. Moreover, an $s_{\pm}$-order which mixes different bands is found at high Hund's coupling, and at high strain anisotropic $s$- and $d$-wave orders emerge. Finally, we show that in a fine-tuned region of parameters a spin-triplet $p$-wave order exists. The combination of strong spin-orbit coupling, interactions, and a sensitivity of the band structure to strain proves a fruitful avenue for engineering new quantum phases.
\end{abstract}

\maketitle
\section{Introduction}
\begin{figure}
\centering    
\includegraphics[width=\linewidth]{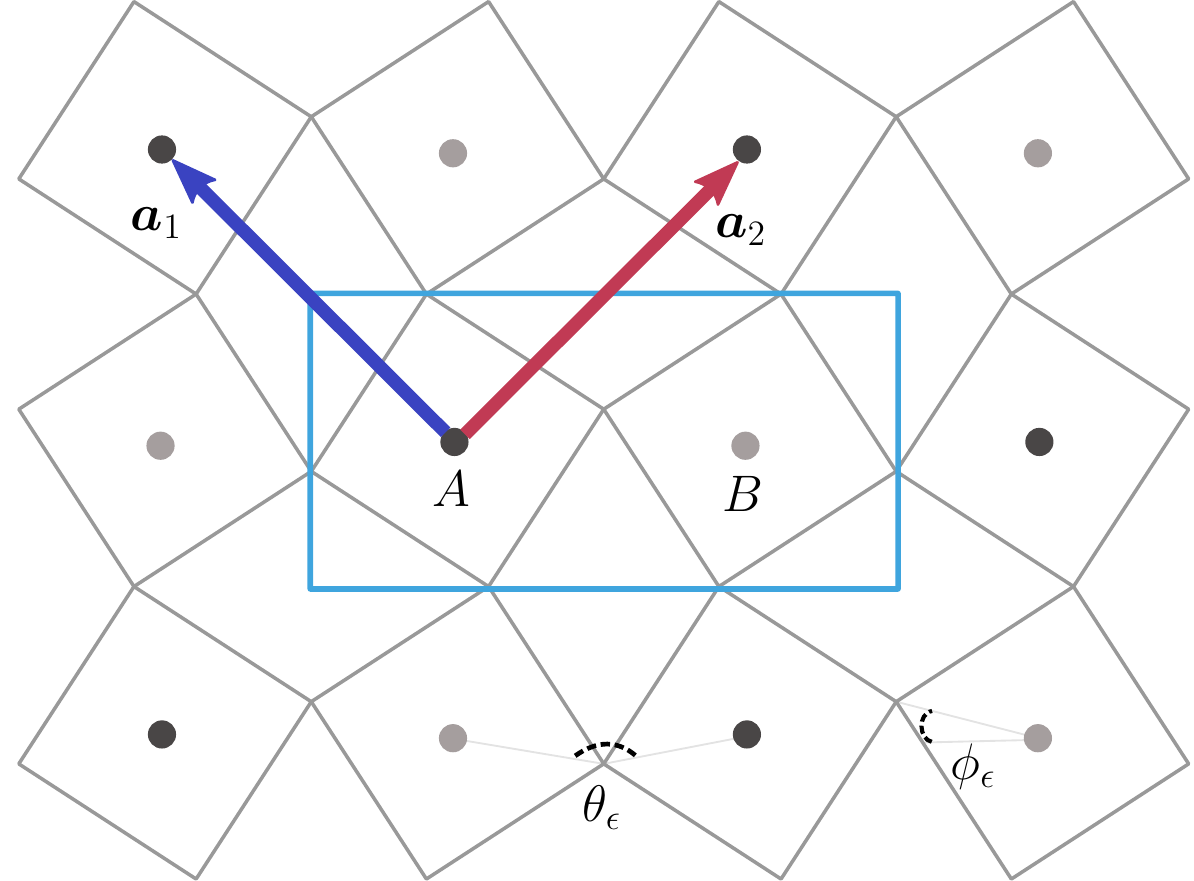}
\caption{\label{fig:canted2siteLattice} Single plane of Sr$_{2}$IrO$_{4}$ where the iridium atoms (gray) form a square lattice, and are surrounded by oxygen octahedra. There are two sites in the unit cell $s=A,B$ with staggered rotations $\phi_{\epsilon}$ of the octahedra, resulting in a bond angle $\theta_\epsilon = 180^{\circ} - 2\phi_{\epsilon} $. The staggered rotation angle increases with compressive strain $\phi_{\epsilon}> \phi_{\epsilon=0}$, with $\phi_{\epsilon=0} \approx 12^{\circ}$.}
\end{figure}

The iridates display a rich phase diagram due to an interplay between strong correlations, spin-orbit coupling, and crystal field effects, all acting on multiple $d$-orbitals which in some cases lead to multiple Fermi surfaces. Various iridates show novel phenomena such as Kitaev and Weyl physics~\cite{Witczak-Krempa2014,Cao2018,Bertinshaw2019}. The first compound in the family of Ruddlesden-Popper perovskite strontium iridates, Sr$_{2}$IrO$_{4}$, consists of stacked quasi-2d layers. In each layer, the iridium atoms form a square lattice, and are surrounded by octahedra of oxygen atoms. As shown in Fig.~\ref{fig:canted2siteLattice}, every other IrO$_{6}$ octahedron is rotated by an angle of $\phi= \pm \phi_{\epsilon=0} \approx \pm 12^{\circ}$ relative to the iridate lattice and we therefore use a two-site basis~\cite{Boseggia2013}. In Sr$_{2}$IrO$_{4}$ the three $t_{2g}$ orbitals are located close to the Fermi surface. However, due to a strong spin-orbit coupling the resulting band structure is better characterized by the on-site total angular momentum $J$ eigenstates, with $J=1/2$ states having $J_{z}=\pm 1/2$ and $J=3/2$ having the projections along the $z$-axis $J_{z}= \pm 1/2, \pm 3/2$. We refer to this basis as the $j$-states. For the undoped compound the electron filling is $n=5$, out of the 6 $t_{2g}$ bands per site, and an antiferromagnetic insulator is found up to $T_{\textrm{N}}=240$~K~\cite{Crawford1994}. For this state, the two bands of mainly $j=1/2$ character are half-filled and located close to the Fermi surface while additional bands of $j=3/2$ character are further from the Fermi level~\cite{Kim2008}.

Experimentally, Sr$_{2}$IrO$_{4}$ shows Fermi arcs and a pseudogap under electron doping~\cite{Kim2014, Cao2016} as well as non-Fermi liquid behavior under hole doping~\cite{Cao2016}. Superconductivity has been predicted for both types of charge doping of Sr$_{2}$IrO$_{4}$ in multiple theoretical works~\cite{Wang2011,Watanabe2013,Yang2014,Meng2014,Gao2015,Nishiguchi2019}. At a first approximation the band structure and the interactions suggest a direct parallel to known high $T_{c}$ superconductors, the cuprates. Indeed, projecting the Hamiltonian on the $j=1/2$ states results in a model similar to those used to describe the cuprates~\cite{Kim2012} and a simplified one-band model therefore predicts $d$-wave superconductivity for electron doped Sr$_{2}$IrO$_{4}$~\cite{Wang2011}.  However, theoretical studies of Sr$_{2}$IrO$_{4}$ found that $d$-wave superconductivity is likely to arise only in a limited range of interaction parameters.  At hole doping, additional pockets of $j=3/2$ character appear at the Fermi level. In this region of the phase diagram, previous studies have predicted Sr$_{2}$IrO$_{4}$ to have either multi-band $s_{\pm}$-wave or a $p$-wave pairing. However, as of yet no superconducting order has been experimentally confirmed for Sr$_{2}$IrO$_{4}$ when chemical doping is the only tuneable parameter~\cite{Kim2016,Chen2015,Battisti2017,Chen2018}. The question then remains whether there could be a tuning parameter that would make superconductivity more favorable. 

The local staggered rotations of iridium sites introduce an in-plane translation symmetry breaking accompanied by additional hybridization of orbitals. These effects have been previously ignored in multi-orbital models of iridate superconductivity~\cite{Lindquist2019}. The rotations increase under compression and as a result the hopping between orbitals at neighboring sites is modified to reflect the new geometry~\cite{Jackeli2009,Boseggia2013,Perkins2014,Liu2015,Torchinsky2015}. Moreover, the orbitals are modified by different amounts such that the bands belonging to the $j=3/2$ subspace move closer to the Fermi surface~\cite{Engstrom2021}. Naturally, the number of Fermi pockets and their orbital composition are important factors for superconductivity. 
As the undoped Sr$_{2}$IrO$_{4}$ is an antiferromagnetic insulator, a prerequisite for any superconducting order is that it must exist in a regime where the system is no longer magnetic. Several experimental studies have shown that by growing Sr$_{2}$IrO$_{4}$ on a substrate with mismatched lattice parameters, the induced compressive epitaxial strain significantly suppresses the magnetic order~\cite{Lupascu2014,Hao2019,Paris2020,Samanta2018,Bhandari2019,Seo2019}. In a variety of known superconductors biaxial, either compressive or tensile, strain has proven to increase the critical temperature~\cite{Trommler2010,Huhne2008,Phan2017,Ahadi2019,Zhang2022} or to induce a superconductivity/insulator phase transition~\cite{Kawasugi2008,Liu2018}. In the current work compressive strain is suggested to induce the same type of phase transition and could thus expand the region of doping where superconductivity can be observed. The purpose of our study is therefore to determine if superconductivity is more likely when strain is applied. More precisely, we aim to answer the following two questions. First, are there regimes of applied strain where a superconducting order is possible? Second, does the in-plane symmetry breaking due to the rotations result in different superconducting orders? 

Experiments in undoped Sr$_{2}$IrO$_{4}$ under high hydrostatic pressure have been performed in recent years~\cite{Haskel2020,Chen2020,Li2021}.  While both hydrostatic pressure and epitaxial strain change the interatomic distances, they do so in different ways.  The experiments approximate the pressure to strain conversion to be $(\Delta a/ a)/\Delta P = -0.146 \%/$GPa~\cite{Haskel2020}.  A transition into a non-magnetic insulating state occurs under a compression of 17GPa. At sufficient pressure beyond that the resistivity shows a rapid increase accompanied by a pressure-induced structural phase transition~\cite{Haskel2020,Chen2020}.  However, it is important to remember that while the effect on in-plane distances is similar, hydrostatic pressure decreases the inter-layer distance while epitaxial strain increases it~\cite{Seo2019}. In the hydrostatic pressure, the $c$-axis compression increases interactions between perovskite layers while the epitaxial strain does not. A persistent insulating state is thus not expected in the realistic regimes of our phase diagrams. In addition, our region of interest is for charge doping, where the insulating nature of the compound is weaker.

Spin-fluctuations are believed to be able to mediate superconductivity in the iridates~\cite{Meng2014, Nishiguchi2019}. Multi-orbital superconductivity has successfully been modeled with spin-fluctuations in other families of materials such as ruthenates~\cite{Romer2019,Kaser2022,Gingras2022} and iron-based superconductors~\cite{Ikeda2010,Schrodi2020}. In this work, a linearized superconducting gap equation (in the static limit) is solved to find regimes where superconductivity is possible. We consider a multi-orbital Hubbard-Kanamori model of Sr$_{2}$IrO$_{4}$ in a rotated two-site basis. The spin susceptibility is calculated via the random phase approximation (RPA). The spin fluctuations are thus dependent on the staggered sublattice rotations. As the rotations increase with an increasing strain and the RPA susceptibility is used to derive the effective particle-particle interaction, the interaction is dependent on the strain. This in turn results in a strain-dependent linearized gap equation for the superconducting order. Magnetic orders can be identified via the RPA susceptibility. We find a large region of strain-induced superconductivity, as well as several possible magnetic orders. The different types of superconducting order are either mediated by spin or pseudospin fluctuations. We find that as the compressive strain is increased the fluctuations become more spin-like in character. Although several types of fluctuations compete in parts of the calculated phase diagrams, the most prevalent type is antiferromagnetic fluctuations in the $j=1/2$ state. These pseudospin fluctuations can mediate a $d$-wave order. Longer range fluctuations in the spin basis instead mediate the $s_{\pm}$-wave superconductivity. For high compressive strain, intraorbital spin fluctuations can become large enough to mediate anisotropic superconducting orders. All superconductivity in the calculated phase diagrams are mediated by fluctuations of spins oriented in-plane. However, there exists regions where ferromagnetic out-of-plane fluctuations are of equal size. The ferromagnetic fluctuations are found to mediate an odd parity $p$-wave order.

The paper is structured as follows. In section~\ref{sec:ModelMethods}, we introduce the underlying tight-binding Hamiltonian of Sr$_{2}$IrO$_{4}$ and how the compressive strain is modeled. In subsection~\ref{sec:SpinFluc}, the model for superconductivity mediated by spin-fluctuations is introduced. The resulting phase diagram is presented in section~\ref{sec:PhaseDiagrams}, for realistic values of model parameters. Additional phase diagrams are shown for a wider variety of possible values for the Hund's and spin-orbit coupling. We then analyze the nature of the magnetic fluctuations in section~\ref{sec:MagOrder}. The types of superconducting orders, and the fluctuations believed to mediate them, are detailed in section~\ref{sec:Supercond}. Finally, in section~\ref{sec:Discussion} we discuss the experimental possibilities of strain-induced superconductivity, and signatures of the found orders. 

\section{Model and methods}\label{sec:ModelMethods}
\subsection{Kinetic Hamiltonian with rotations}\label{sec:KinHam}
The band structure of Sr$_{2}$IrO$_{4}$ can be modeled by a spin-orbit coupled tight-binding Hamiltonian:
\begin{equation}\label{eq:HnonInt}
H= H_{\textrm{kin}} + H_{\textrm{SOC}}
\end{equation}
We consider a 2-site orbital-spin basis: $\boldsymbol{c}=$ $( c_{\boldsymbol{k},A, yz, \uparrow},$ $ c_{\boldsymbol{k},A, yz, \downarrow},$ $c_{\boldsymbol{k},A, xz, \uparrow},$ $c_{\boldsymbol{k},A, xz, \downarrow},$ $c_{\boldsymbol{k},A, xy, \uparrow},$ $c_{\boldsymbol{k},A, xy, \downarrow},$ $c_{\boldsymbol{k},B, yz, \uparrow},$ $c_{\boldsymbol{k},B, yz, \downarrow},$ $c_{\boldsymbol{k},B, xz, \uparrow},$ $c_{\boldsymbol{k},B, xz, \downarrow},$ $c_{\boldsymbol{k},B, xy, \uparrow},$ $c_{\boldsymbol{k},B, xy, \downarrow})$. For each spin $\sigma=\uparrow, \downarrow$ the kinetic terms have intra- and inter-sublattice hopping:
\begin{equation}
H_{\textrm{kin}} = \left( \begin{array}{c c }
H_{AA} & e^{i k_{x}} H_{AB} \\
e^{-i k_{x}} H_{AB}^{\dagger} & H_{BB}
\end{array} \right)
\end{equation}
\begin{equation}
H_{AA} = \left( \begin{array}{c c c}
\epsilon_{d} & \epsilon_{1d} & 0 \\
\epsilon_{1d} & \epsilon_{d} & 0 \\
0 & 0 & \epsilon^{xy}_{d} 
\end{array} \right), H_{AB} =  \left( \begin{array}{c c c}
 \epsilon_{yz} & - \epsilon_{rot}&0 \\
 \epsilon_{rot}& \epsilon_{xz} & 0 \\
 0 & 0 & \epsilon_{xy}
\end{array} \right)
\end{equation}
and $H_{BB}=H_{AA}$. The factor $e^{i k_{x}}$ arises from the choice of unit cell, where the two sublattice sites are chosen as in Fig.~\ref{fig:canted2siteLattice} and the lattice spacing, $a$, is set to $1$. The hopping terms are
\begin{equation}\label{eq:kinTerms}
\begin{aligned}
\epsilon_{xy}&= 2 t \left( \cos k_{x} + \cos k_{y} \right)\\
 \epsilon_{yz}&= 2 \left( t_{\delta} \cos k_{x} + t_{1} \cos k_{y} \right) \\
\epsilon_{xz}&= 2 \left( t_{1} \cos k_{x} + t_{\delta} \cos k_{y} \right)\\ 
 \epsilon_{rot}&= 2 t' \left( \cos k_{x} + \cos k_{y} \right)\\
\epsilon^{xy}_{d}& = 4 t_{n} \cos k_{x} \cos k_{y} +\mu_{xy}\\
\epsilon_{1d}&= 4 t_{1d} \sin k_{x} \sin k_{y} \\
\epsilon_{d}&= 4 t_{nd} \cos k_{x} \cos k_{y}.
\end{aligned}
\end{equation}
The hopping values are $(t, t_{1}, t_{\delta}, t',$ $ t_{n}, t_{1d}, t_{nd}, \mu_{xy})=$ $-0.36(1, 0.882, 0.260,0.199, 0.559, $ $0.019, -0.010, 0.7)$eV, when no strain is applied. The staggered rotations result in the non-zero inter-orbital hopping, $\epsilon_{rot}$ and $\epsilon_{1d}$, between the $yz$- and $xz$-orbitals. When the compressive strain, $\epsilon<0$, is increased the hopping parameters change. We use a linear strain dependence as in Ref.~\cite{Engstrom2021}, following data by Ref.~\cite{Seo2019}:
\begin{equation} \label{eqn:strainT}
\begin{aligned}
t (\epsilon)&=  t \left(1  + \rho \epsilon \right)\\
t_{1} (\epsilon)&= t_{1} \left(1  + \rho_{1} \epsilon \right) \\
t' (\epsilon)&= t' \left(1  + \rho' \epsilon \right) \\
t_{n} (\epsilon)&=  t_{n} \left(1  + \rho_{n} \epsilon \right)\\
t_{\delta} (\epsilon)&=  t_{\delta} \left(1  + \rho_{\delta} \epsilon \right)\\
t_{1d} (\epsilon)&=  t_{1d} \left(1  + \rho_{1d} \epsilon \right)\\
t_{nd} (\epsilon)&=  t_{nd} \left(1  + \rho_{nd} \epsilon \right)\\
\phi_{\epsilon}&= \phi_{\epsilon=0} \left(1  + \rho_{\phi} \epsilon \right) 
\end{aligned}
\end{equation}
with $(\rho,\rho_{1}, \rho',  \rho_{n}, \rho_{\delta}, \rho_{1d}, \rho_{nd},  \rho_{\phi} ) =$ (0.014,-0.251, -0.309,   -0.048,  0, 0,-0.02,-0.085). $\phi_{\epsilon}$ is angle of the staggered rotations. The approximation considers not only rigid rotations of the oxygen octahedra but also changes in bond lengths, as was recently proposed to be a more accurate description of the strain effect in Ref.~\cite{Parschke2022}. The strain, and the associated rotations of the octahedra, increase the inter-orbital $t'$ and intra-orbital $t_{1}$. On the other hand hopping within the $xy$-orbital decreases. The effect of strain on the tetragonal splitting $\mu_{xy}$ is not discussed here and is deferred to Appendix~\ref{sec:AppC}. 
\begin{figure}
\centering   
\includegraphics[width=\linewidth]{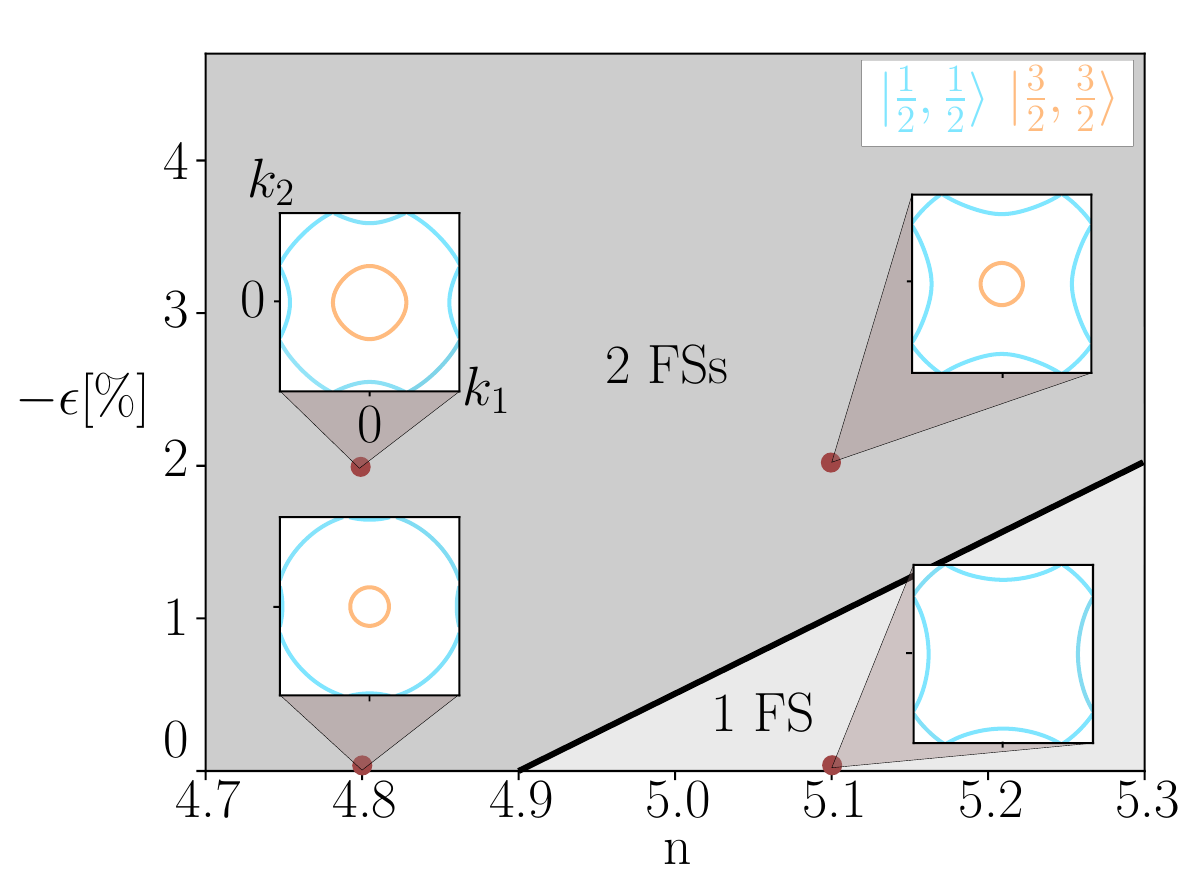}  
\caption{\label{fig:FS_n_strain}The Fermi surface (FS) or surfaces are shown as a function of strain and doping. There are one or two types of pockets whose spin and orbital character is indicated by color. These are found by diagonalizing the non-interacting model in Eq.~\eqref{eq:HnonInt}. The number of pockets and a few examples of the Fermi surface are shown for $\lambda=0.6$eV. The pocket of $j=3/2$ character is only present at hole doping $n \leq 4.9$ at $\epsilon=0$. For increasing compressive strain two types of pockets are present for all doping. }
\end{figure}
It has been well-established that the spin-orbit coupling in Sr$_{2}$IrO$_{4}$ is large enough for each band to have a clear character of either total angular momenta $j=1/2$ or $j=3/2$. The atomic SOC is
\begin{equation}
H_{\textrm{SOC}} = \frac{\lambda}{2} \sum_{\alpha \beta, \sigma \sigma'} \sum_{s=A,B} \boldsymbol{L}_{\alpha \beta} \cdot\boldsymbol{\sigma}_{\sigma \sigma'} c^{\dagger}_{ \boldsymbol{k} s \alpha \sigma} c_{\boldsymbol{k} s \beta \sigma'}
\end{equation}
where $\boldsymbol{\sigma}=  \left( \sigma^{x},\sigma^{y},\sigma^{z} \right)$ are the Pauli matrices in the spin basis $\sigma=\uparrow, \downarrow$ with 
respect to the $z$-direction, and
\begin{equation}
\boldsymbol{L} = \left( \begin{bmatrix}
 0 & 0 & 0 \\
 0 & 0 & -i \\
 0 & i& 0
\end{bmatrix}, \begin{bmatrix}
 0 & 0 & i \\
 0 & 0 & 0 \\
 -i & 0& 0
\end{bmatrix},\begin{bmatrix}
 0 & -i & 0 \\
 i & 0 & 0 \\
 0 & 0& 0
\end{bmatrix} \right).
\end{equation}
The eigenstates of $H_{\textrm{SOC}}$ are the $j$-states, with associated annihilation operators: $\boldsymbol{a}_{m, \tau} = \sum_{\alpha, \sigma} M_{(m, \tau),(\alpha, \sigma)}\boldsymbol{c}_{\alpha, \sigma}$. Where
\begin{equation} \label{eq:Mmat}
M = \left(
\begin{array}{cccccc}
 0 & \frac{1}{\sqrt{3}} & 0 & -\frac{i}{\sqrt{3}} & \frac{1}{\sqrt{3}} & 0 \\
 \frac{1}{\sqrt{3}} & 0 & \frac{i}{\sqrt{3}} & 0 & 0 & -\frac{1}{\sqrt{3}} \\
 0 & \frac{1}{\sqrt{6}} & 0 & -\frac{i}{\sqrt{6}} & -\sqrt{\frac{2}{3}} & 0 \\
- \frac{1}{\sqrt{6}} & 0 & -\frac{i}{\sqrt{6}} & 0 & 0 & -\sqrt{\frac{2}{3}} \\
 0 & -\frac{1}{\sqrt{2}} & 0 & -\frac{i}{\sqrt{2}} & 0 & 0 \\
 \frac{1}{\sqrt{2}} & 0 & -\frac{i}{\sqrt{2}} & 0 & 0 & 0
\end{array} 
\right) 
\end{equation}
with $\boldsymbol{c}_{\alpha, \sigma}=( c_{yz, \uparrow},  c_{yz, \downarrow},  c_{xz, \uparrow},  c_{xz, \downarrow}, c_{xy, \uparrow} , c_{xy, \downarrow}) $ and $\boldsymbol{a}_{m, \tau} =( a_{1, +} , a_{1, -} , a_{2, +}, a_{2, -} , a_{3, +}, a_{3, -} )$. In this basis each site has the states $a_{m, \tau}$, where $m$ denotes the total angular momentum and its $z$-axis projection $\left( j, j_{z} \right)$ such that $1=\left( \frac{1}{2}, \pm \frac{1}{2} \right), 2=\left( \frac{3}{2}, \pm \frac{1}{2} \right), 3=\left( \frac{3}{2}, \pm \frac{3}{2} \right)$ and the projections along the $z$-axis are labeled by $\tau=\pm$. The projections $\tau$ can be treated as pseudospins, here not mixing sublattice and spin degrees of freedom but orbital and spin~\cite{Sigrist1991}. The total of 12 bands $b_{\boldsymbol{k}, n}$ have eigenvalues $\xi_{\boldsymbol{k}, n}$ and are connected to the orbital basis via $c_{\boldsymbol{k}, j}  = \sum_{n} U_{\boldsymbol{k}, j n} b_{\boldsymbol{k}, n}$. The only spin-mixing in the Hamiltonian comes from the atomic spin-orbit coupling and all hopping terms are pseudospin-conserving. The non-interacting Hamiltonian is therefore separable into pseudospin $\tau=+,-$ sectors, containing the states $ \{(yz, \downarrow), (xz, \downarrow), (xy, \uparrow) \}$ and $\{(yz, \uparrow), (xz, \uparrow), (xy, \downarrow) \}$ respectively. Therefore the 12 bands can be described by 6 bands in each pseudospin-sector $n_{\tau}$. 

As the strain modifies the hopping parameters, both the shape and the number of pockets at the Fermi surface changes. Therefore the Fermi surface is different at every point in the phase diagrams. As can be seen in Figs.~\ref{fig:FS_n_strain}, there are pockets belonging to the bands of $\left( j, j_{z} \right) = \left( \frac{1}{2}, \pm \frac{1}{2} \right)$ present at every point in the phase diagram. For fillings corresponding to hole doping, $n<5$, another pocket with $\left( j, j_{z} \right) = \left( \frac{3}{2}, \pm \frac{3}{2} \right)$ appears around $(k_{1}, k_{2}) =(0,0)$. The strain increases the bandwidth and cause the $(\frac{3}{2},\pm \frac{3}{2})$ pocket to appear for all doping values. As shown further in Appendix~\ref{sec:AppB} the size of the $j=1/2$ electron pocket increases with strain.
 
\subsection{Spin-fluctuation mediated superconductivity}\label{sec:SpinFluc}
To analyze the possibility of spin fluctuation mediated superconductivity, we solve a linearized gap equation in the static limit and normal state. A similar calculation has been performed previously for Sr$_{2}$IrO$_{4}$ in Ref.~\cite{Nishiguchi2019}, for a model without staggered rotations or strain. In general, the particle-hole and particle-particle self-energies are defined by the Dyson-Gorkov equation:
\begin{equation}\label{eq:DysonGorkov}
\boldsymbol{G}(k) = \boldsymbol{G}^{(0)}(k) + \boldsymbol{G}^{(0)}(k) \boldsymbol{\Sigma}(k) \boldsymbol{G}(k)
\end{equation}
with
\begin{equation}
\displaystyle \boldsymbol{G}(k) = \left[ \begin{array}{c c} G(k) & F(k) \\
\bar{F}(k) & \bar{G}(k) \end{array} \right], \qquad \boldsymbol{\Sigma}(k) = \left[ \begin{array}{c c} \Sigma (k) & \Delta (k)\\
\bar{\Delta} (k) & \bar{\Sigma} (k) \end{array} \right]
\end{equation}
\begin{equation}
\displaystyle  \boldsymbol{G}^{(0)}(k) = \left[ \begin{array}{c c} G^{(0)}(k) & 0 \\
0 &  \bar{G}^{(0)}(k)
\end{array} \right]
\end{equation}
where $G(k)$ is the particle-hole and $\bar{G}(k)$ the hole-particle Green's functions 
\begin{equation}
    G_{ab}(k) = - \int_{0}^{\beta} d \tau e^{i \omega_{n} \tau} \langle T_{\tau} c_{\boldsymbol{k} a}(\tau) c_{\boldsymbol{k} b}^{\dagger} (0) \rangle
\end{equation}
\begin{equation}
    \bar{G}_{ab}(k) = - \int_{0}^{\beta} d \tau e^{i \omega_{n} \tau} \langle T_{\tau} c_{-\boldsymbol{k} a}^{\dagger}(\tau) c_{-\boldsymbol{k} b} (0) \rangle
\end{equation}
for crystal momentum $\boldsymbol{k}$ and fermionic Matsubara frequencies $\omega_{n}= (2n+1) \pi /\beta$, with $\beta= \hbar/(k_{\textrm{B}}T)$. In the orbital-spin-sublattice basis $a=(\alpha, \sigma, s)$ with orbital $\alpha$, spin $\sigma$, and sublattice $s$. The non-interacting Green's functions are calculated with the operators $c_{\boldsymbol{k} a}$ for the Hamiltonian in Eq. \eqref{eq:HnonInt}. In the band ($n$) and orbital-spin-sublattice basis respectively:
\begin{equation} \label{eqn:G0}
G^{(0)}_{ n}(\boldsymbol{k}, i \omega_{n}) = \left[ i \omega_{n} - \xi_{\boldsymbol{k}, n} \right]^{-1}
\end{equation}
\begin{equation}
 G^{(0)}_{ab}(k) = \sum_{n} U_{\boldsymbol{k}, a n} G^{(0)}_{ n}(k) U^{\dagger}_{\boldsymbol{k}, n b }.
\end{equation}
For the non-interacting model $G^{(0)}_{ab }(k)= -\bar{G}^{(0)}_{ba }(-k)$. Under the approximations of this work, only the particle-particle self-energy is calculated. In a multi-orbital system, the non-interacting particle-hole susceptibility $\hat{\chi}_{0}(k)= \hat{\chi}_{0}^{ph}(k)$ is a tensor calculated from the non-interacting Green's functions in the $N_{k} \times N_{k}$-lattice we are considering:
\begin{equation}
\chi^{ph}_{0,abcd}(q) = \frac{1}{N_{k}^{2} \beta} \sum_{k} G^{(0)}_{ac}(q+k) \bar{G}^{(0)}_{ bd }(-k)
\end{equation}
Using the well-known summation for the Lindhard function over all fermionic Matsubara frequencies $\omega_{n}$ and the analytic continuation:
\begin{equation}\label{eq:chi0}
\begin{aligned}
&\chi_{0, abcd}(\boldsymbol{q}, i \omega_{n}\rightarrow i \delta) = \\ &\frac{1}{N_{k}^{2} } \sum_{\boldsymbol{k},n,n'} \left[ U_{\boldsymbol{k}+ \boldsymbol{q}}\right]_{a n} \left[ U^{\dagger}_{\boldsymbol{k}+\boldsymbol{q}}\right]_{n c} \left[  U_{\boldsymbol{k}}\right]_{d n'}  \left[ U^{\dagger}_{\boldsymbol{k}}\right]_{ n' b}  \\ &\times \frac{f(\xi_{\boldsymbol{k}+ \boldsymbol{q},n}, T) - f(\xi_{\boldsymbol{k},n'}, T) }{i \delta -\left( \xi_{\boldsymbol{k}+ \boldsymbol{q},n} - \xi_{\boldsymbol{k},n'} \right)}
\end{aligned}
\end{equation}
where $\delta$ is a small number, here set to $10^{-4}$eV. $f(\xi_{\boldsymbol{k},n}, T)$ is the Fermi-Dirac distribution at temperature $T$. The tensor $\hat{\chi}_{0}(q)$ can be written as a rank 4 tensor. The number of spins $N_{\sigma}=2$, orbitals $N_{o}=3$, and sublattices $N_{s}=2$, results in a tensor of dimension $12 \times 12 \times 12 \times 12$. However, to separate the spin degrees of freedom we can reshape the tensor to have the dimension $N_{\sigma}^{2} \times N_{\sigma}^{2} \times (N_{o}  N_{s})^{2} \times (N_{o}  N_{s})^{2}$. In terms of spin and orbital indices, the basis for the tensor on this form is given as all combinations of two indices $(\sigma \sigma') \times (\alpha_{s} \beta_{s'})$. The spin combinations are $(\sigma \sigma')=\{(\uparrow \uparrow),(\uparrow \downarrow), (\downarrow \uparrow), (\downarrow \downarrow) \}$, and $(\alpha_{s} \beta_{s'})$ are the 36 combinations of orbitals $\alpha=yz,xz, xy$ and sublattices $s=A,B$.

Interactions are treated by calculating the irreducible particle-particle vertex, which is modified by spin-fluctuations from the random phase approximation (RPA). The irreducible vertex is given by the parquet equations~\cite{Bickers2004}, where here the bare vertex is $\hat{\Gamma}_{0}= \hat{V}$. The bare vertex is the Hubbard-Kanamori interaction~\cite{Kanamori1963}, which is defined in real space as
\begin{equation}
\begin{aligned}
H_{\textrm{I}}= &\displaystyle U \sum_{\boldsymbol{j}, \alpha} n_{\boldsymbol{j} \alpha \uparrow} n_{\boldsymbol{j} \alpha \downarrow} \\ &
+ \displaystyle \sum_{\boldsymbol{j}, \alpha \neq \beta} J_{\textrm{H}} \left[ c^{\dagger}_{\boldsymbol{j} \alpha \uparrow} c^{\dagger}_{\boldsymbol{j} \beta \downarrow} c_{\boldsymbol{j} \alpha \downarrow} c_{\boldsymbol{j} \beta \uparrow} + c^{\dagger}_{\boldsymbol{j} \alpha \uparrow} c^{\dagger}_{\boldsymbol{j} \alpha \downarrow} c_{\boldsymbol{j} \beta \downarrow} c_{\boldsymbol{j} \beta \uparrow} \right] \\ &
+ \displaystyle \sum_{\boldsymbol{j}, \alpha < \beta, \sigma} \left[ U' n_{\boldsymbol{j} \alpha \sigma} n_{\boldsymbol{j} \beta \bar{\sigma}} + \left(U' - J_{\textrm{H}} \right) n_{\boldsymbol{j} \alpha \sigma} n_{\boldsymbol{j} \beta \sigma} \right]
\end{aligned}
\end{equation}
with the intraorbital Hubbard interaction $U$, the Hund's coupling $J_{\textrm{H}}$, and the interorbital repulsion $U' = U- 2 J_{\textrm{H}}$. Following the notation of Ref.~\cite{Nishiguchi2019}, we define the bare vertex $\hat{V}$ as:
\begin{equation}
H_{\textrm{I}}= \displaystyle \frac{1}{N^{2}_{k}} \sum_{\boldsymbol{k}, \boldsymbol{q}} \sum_{a,b,c,d} V_{abcd} (\boldsymbol{q}) c^{\dagger}_{\boldsymbol{k}, a} c^{\dagger}_{-\boldsymbol{k}, c} c_{-(\boldsymbol{k}-\boldsymbol{q}), b} c_{\boldsymbol{k}-\boldsymbol{q}, d}
\end{equation}
for all orbit-spin-sublattice indices $a=(\alpha, \sigma, s)$. On the same rank 4 tensor structure form as the susceptibility, the tensor $\hat{V}$ can be divided into spin sectors
\begin{equation} \label{eq:Vbare}
\hat{V} = \left( \begin{array}{c c c c}
 \hat{V}^{ \uparrow \uparrow \uparrow \uparrow} & 0 & 0  & \hat{V}^{ \uparrow \uparrow \downarrow \downarrow} \\
 0 & \hat{V}^{\uparrow \downarrow \uparrow \downarrow} & 0 & 0 \\
 0 & 0 & \hat{V}^{ \downarrow \uparrow \downarrow \uparrow}  & 0\\
 \hat{V}^{ \downarrow \downarrow \uparrow \uparrow} &0 & 0& \hat{V}^{ \downarrow \downarrow \downarrow \downarrow}
\end{array} \right)
\end{equation}
with only on-site terms $s=s'$. Each matrix $\hat{V}^{ \sigma_{1} \sigma_{2} \sigma_{3} \sigma_{4} }$ in the orbit-sublattice basis is defined as
\begin{equation}
V^{\uparrow \uparrow \uparrow \uparrow}_{\alpha_{s}\beta_{s}\gamma_{s}\delta_{s}}= V^{\downarrow \downarrow \downarrow \downarrow}_{\alpha_{s}\beta_{s}\gamma_{s}\delta_{s}} = \begin{cases} U'-J_{\textrm{H}} & (\alpha=\gamma \neq \beta=\delta)\\
-U'+J_{\textrm{H}} & (\alpha=\beta \neq \gamma=\delta) \\ 
0 & (\textrm{otherwise})
 \end{cases}
\end{equation}
\begin{equation}
V^{\uparrow \uparrow \downarrow \downarrow}_{\alpha_{s}\beta_{s}\gamma_{s}\delta_{s}}= V^{\downarrow \downarrow \uparrow \uparrow}_{\alpha_{s}\beta_{s}\gamma_{s}\delta_{s}} = \begin{cases} -U & (\alpha=\beta = \gamma= \delta) \\
-J_{\textrm{H}} & (\alpha=\gamma \neq \beta=\delta) \\ 
-U' & (\alpha=\beta \neq \gamma=\delta) \\ 
-J_{\textrm{H}} & (\alpha=\delta \neq \beta=\gamma) \\ 
0 & (\textrm{otherwise})
 \end{cases}
\end{equation}
\begin{equation}
V^{\uparrow \downarrow \uparrow \downarrow}_{\alpha_{s}\beta_{s}\gamma_{s}\delta_{s}}= V^{\downarrow \uparrow \downarrow \uparrow}_{\alpha_{s}\beta_{s}\gamma_{s}\delta_{s}} = \begin{cases} U & (\alpha=\beta = \gamma= \delta) \\
U' & (\alpha=\gamma \neq \beta=\delta) \\ 
J_{\textrm{H}} & (\alpha=\beta \neq \gamma=\delta) \\ 
J_{\textrm{H}} & (\alpha=\delta \neq \beta=\gamma) \\ 
0 & (\textrm{otherwise})
 \end{cases}
\end{equation}
for each sublattice $s=A,B$ and for orbital indices $\alpha, \beta, \gamma, \delta$. The effective particle-particle vertex~\cite{Bickers2004} is:
\begin{equation}\label{eq:VppEff}
\hat{\Gamma}^{pp}(q= K-K')=- \frac{1}{2} \hat{V} - \hat{V} \hat{\chi}(K-K')\hat{V}
\end{equation} 
The susceptibility $\hat{\chi}(q)$ is approximated as the RPA susceptibility:
\begin{equation}\label{eq:chiRPA}
\hat{\chi}(q) = \left( 1 - \hat{\chi}_{0}(q) \hat{V} \right)^{-1} \hat{\chi}_{0}(q)
\end{equation}
which uses the same bare vertex $\hat{V}$ and has the same tensor structure as $\hat{\chi}_{0}(q)$. The tensor multiplication is defined as a matrix multiplication in the spin and orbital-sublattice combinations
\begin{equation}\label{sec:tensorMul}
  \left[ \hat{A} \hat{B} \right]^{(\sigma \sigma')_{1}(\sigma \sigma')_{2}}_{(\alpha_{s} \beta_{s'})_{1}(\alpha_{s} \beta_{s'})_{2}}  = 
\sum_{\tilde{\alpha}, \tilde{\beta}, \tilde{\sigma}, \tilde{\sigma}'} A^{(\sigma \sigma')_{1} (\tilde{\sigma} \tilde{\sigma}')}_{(\alpha_{s} \beta_{s'})_{1} (\tilde{\alpha} \tilde{\beta})} B^{(\tilde{\sigma} \tilde{\sigma}')(\sigma \sigma')_{2}}_{(\tilde{\alpha} \tilde{\beta}) (\alpha_{s} \beta_{s'})_{2}}
\end{equation}
The effective vertex is not further spin-diagonalized into spin-singlet and spin-triplet vertices as our regime of intermediate to strong spin-orbit coupling inherently mixes the two sectors. The linearized gap equation can be obtained from the Luttinger-Ward functional~\cite{Baym1961,Baym1962}, where the linearization of the anomalous Green's function is given by the Dyson-Gorkov equations Eq.~\eqref{eq:DysonGorkov}. 
\begin{equation}\label{eq:linGap}
\Delta_{ab}(k) = \frac{1}{N \beta} \sum_{k'} \sum_{a' b'}  \Gamma^{pp}_{a a' b' b}(k- k') F_{a' b'}(k')
\end{equation}
\begin{equation}
F_{a' b'}(k') = \sum_{\mu \nu} G_{a' \mu}(k') \bar{G}_{\nu b'}(k') \Delta_{\mu \nu}(k') 
\end{equation}
where $\beta$ is the inverse temperature and $N=N_{k}^{2}$. Each index here runs over all orbit-spin-sublattice combinations $a=(\alpha, \sigma, s)$. Further, we apply both the static ($\omega_{n} \rightarrow \delta = 10^{-4}$eV, $\hat{\Delta}(k) \rightarrow \hat{\Delta}(\boldsymbol{k})$, $\Gamma^{pp}(q) \rightarrow \Gamma^{pp}(\boldsymbol{q})$) and normal state ($\hat{G}(k') \rightarrow \hat{G}^{(0)}(k')$) approximations. The linearized gap equation can be solved as an eigenvalue problem, as a version of the Eliashberg equation:
\begin{equation}\label{eq:SCeq}
\lambda_{e}\Delta_{a b}(\boldsymbol{k}) =\frac{1}{N} \sum_{\boldsymbol{k}',a' b',\mu \nu} \Gamma^{pp}_{ a a' b' b}(\boldsymbol{k}- \boldsymbol{k}') \phi_{a' b' \mu \nu}^{\boldsymbol{k}'}\Delta_{\mu \nu} (\boldsymbol{k}') 
\end{equation}
with
\begin{equation}
\phi_{a' b' \mu \nu}^{\boldsymbol{k}'}=-\frac{1}{\beta}\sum_{\omega_{n}} G^{(0)}_{a' \mu}(k') \bar{G}^{(0)}_{\nu b'}(k') 
\end{equation}
A largest eigenvalue of unity or higher $\lambda_{e} \geq 1$ indicates a possible superconducting order. A non-explicit summation over Matsubara frequencies is performed, like in Eq.~\eqref{eq:chi0}, and the equation depends only on momentum $\boldsymbol{k}$.

While this system is strongly interacting one might expect a multitude of order parameters including magnetic orders. We use the Stoner criterion to identify phases in the particle-hole channel. Methods attempting to treat particle-hole and particle-particle self-energies on equal footing are beyond the scope of this work~\cite{Bickers1989, Bickers1991}. The RPA susceptibility, Eq.~\eqref{eq:chiRPA}, will pass a critical point and diverge, when the Hartree-Fock term in the particle-hole channel has an eigenvalue above unity at any $k$-point which we name $Q$~\cite{Bickers1989, Bickers1991}:
\begin{equation}\label{eq:StonerDef}
\textrm{max eig} \left[ \hat{\chi}_{0}(Q) \hat{V} \right] = 1
\end{equation}
The particle-hole instability is given by the eigenvector to the tensor $\hat{\chi}_{0}(Q) \hat{V}$, which can be unfolded into a matrix. A rank 4 tensor $C_{ijkl}$ of dimension $N_{1} \times N_{2} \times N_{1} \times N_{2} $ can be mapped onto a matrix $C_{\mu \nu}$ of dimension $N_{1} N_{2} \times N_{1} N_{2}$ via
\begin{equation}
    \begin{aligned}
        \mu = i + N_{1}(j-1), & \qquad \mu \in \left[1, \dots, N_{1} N_{2}\right] \\
        i = 1 + \textrm{mod}(\mu -1 , N_{1}), & \qquad i \in \left[1, \dots, N_{1} \right] \\
        j = 1 + \textrm{div}(\mu -1 , N_{1}), & \qquad j \in \left[1, \dots, N_{2}\right]
    \end{aligned}
\end{equation}
where "mod" is the modulus operation and "div" is integer division. The given mapping preserves the defined tensor multiplication, in Eq.~\eqref{sec:tensorMul}, as matrix multiplication in the unfolded matrix. The type of instability can also be classified by the magnetic channel it occurs in. The susceptibility can be spin block-diagonalized by rewriting it in the basis of magnetic operators $m^{z}_{\tilde{\alpha} \tilde{\beta}}= \frac{1}{\sqrt{2}} (c^{\dagger}_{\tilde{\alpha} \uparrow } c_{\tilde{\beta} \uparrow } -c^{\dagger}_{\tilde{\alpha} \downarrow } c_{\tilde{\beta} \downarrow })$: 
\begin{equation}
    \hat{\chi}^{z}_{\tilde{\alpha} \tilde{\beta} \tilde{\gamma} \tilde{\delta}}(q) = \frac{1}{N_{k}^{2}} \int_{0}^{\beta} d \tau e^{i \omega_{n} \tau} \langle T_{\tau} m^{z}_{\tilde{\alpha} \tilde{\beta}, \boldsymbol{q}}(\tau) m^{z}_{ \tilde{\gamma} \tilde{\delta}, -\boldsymbol{q}}(0) \rangle_{c}
\end{equation}
with the indices $\tilde{\alpha} = \alpha_{s}$ running over all orbital $\alpha=yz,xz,xy$ and sublattice $s=A,B$ combinations. The magnetic channels are
\begin{equation}\label{eq:chiz}
\hat{\chi}^{z}(q) = \frac{1}{2} \left( \hat{\chi}^{\uparrow \uparrow \uparrow \uparrow}(q) - \hat{\chi}^{\uparrow \uparrow \downarrow \downarrow}(q) \right)
\end{equation}
\begin{equation}\label{eq:chiinPlane}
\hat{\chi}^{+}(q) =  \hat{\chi}^{\uparrow \downarrow \uparrow \downarrow}(q), \qquad \hat{\chi}^{-}(q) =  \hat{\chi}^{ \downarrow \uparrow \downarrow \uparrow}(q)
\end{equation}
with the out-of-plane spin $\hat{\chi}^{z}(q)$ and in-plane spin $\hat{\chi}^{\pm}(q)$ channels \footnote{The spin block-diagonalization introduces an additional density channel $\hat{\chi}^{d}(q)$. No density-channel instability is found in this work and the peaks are significantly smaller than in the magnetic channels. This remains true for $\hat{\chi}^{d}_{J}(q)$ in the $j$-state basis.}. As the $j$-state basis describes the bands better than the orbital-spin basis, the $\chi(q)$-terms can be transformed, via Eq.~\eqref{eq:Mmat}, as
\begin{equation}\label{eq:chiJtransf}
\chi_{J, ijkl}(q) = \sum_{\alpha \beta \gamma \delta}  M_{i \alpha} M^{\dagger}_{\gamma k} M_{l \delta} M^{\dagger}_{ \beta j} \chi_{\alpha \beta \gamma \delta}(q)
\end{equation}
where $M$ is the transformation from spin and orbit to the total angular momentum basis. In this basis, the susceptibility is similarly divided into different pseudospin channels $\hat{\chi}^{z}_{J}(q)$ and $\hat{\chi}^{\pm}_{J}(q)$.

\subsection{Computational details}
All RPA calculations, for the susceptibility Eq.~\eqref{eq:chiRPA} and superconductivity Eq.~\eqref{eq:SCeq} were performed on a $N_{k} \times N_{k}=46 \times 46$ momentum lattice. The finite momentum resolution limits the lowest accessible temperature. Peaks in the susceptibility cannot be narrower than the lattice spacing and we thus require some thermal broadening to get reliable results. In this work we use $k_{\textrm{B}}T=0.001$eV ($T \approx 11$K). The temperature was chosen such that less than a $10\%$ change in largest eigenvalue was found when going from a lattice of size $N_{k}=32$ to $N_{k}=46$, at most points. In the data of Figs.~\ref{fig:PDchiRest}b \& \ref{fig:maxPeakspwave} a larger change is observed, therefore calculations for these figures where done at $N_{k}=64$. The specific temperature is of interest for a potential $d$-wave superconducting order in the electron doped compound. The 2016 experiment in Ref.~\cite{Kim2016} observed an order with this symmetry below $T \approx 30$K, with a maximum at $T \approx 10$K. The largest value of the linearized gap equation is found via the Arnoldi method, with a convergence criterion of $10^{-7}$.
\begin{figure}
\centering    
\includegraphics[width=0.48\textwidth]{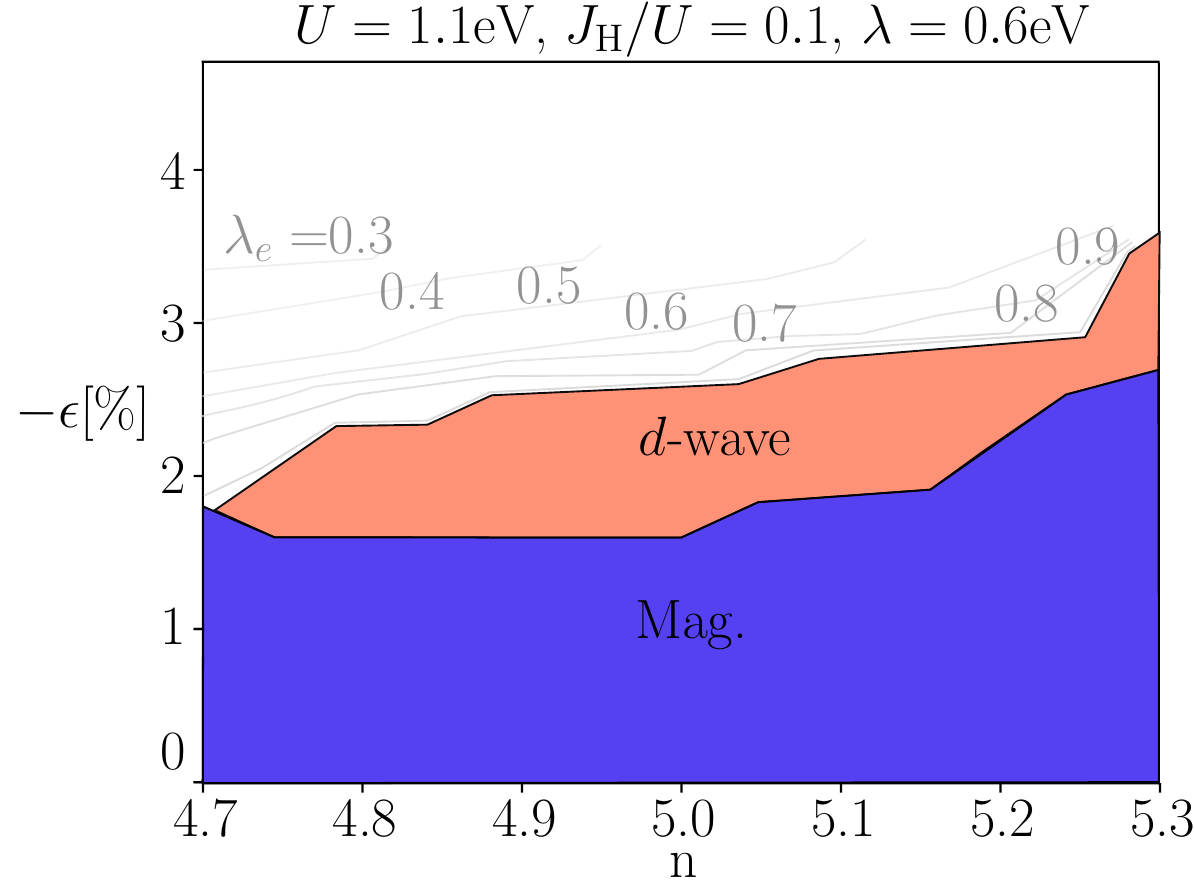} 
\caption{\label{fig:PDr}The phase diagram for charge doping and compressive strain are shown at $U=1.1$eV$\approx 3 |t|$ for realistic values of the spin orbit $\lambda$ and Hund's couplings $J_{\text{H}}$. Two types of regions are found in the RPA calculations: the magnetic region where the Stoner criterion has been met and a superconducting order with $d$-wave symmetry. The nature of the magnetic transition is characterized in Fig.~\ref{fig:PDchiJ12}. In the normal state the largest eigenvalue $\lambda_{e}<1$, and the contours of the values are shown up to $\epsilon=-3.5\%$. The phase diagram extends to the higher strain values as to be comparable to Fig.~\ref{fig:PDHund}, in this regime no superconducting order is possible.}
\end{figure}

\begin{figure*}
\centering 
\includegraphics[width=\textwidth]{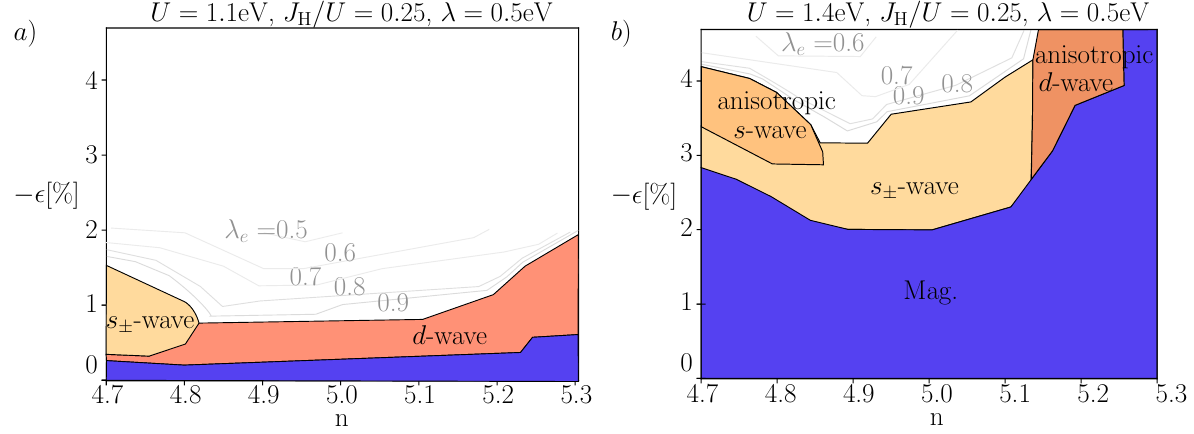} 
\caption{\label{fig:PDHund}The phase diagrams for for a lower SOC and a higher Hund's, at a) $U=1.1$eV$\approx 3 |t|$ and b) $U=1.4$eV$\approx 4|t|$. In a) there are two types of superconducting orders: the $d$-wave and another with $s_{\pm}$-wave symmetry. In b) there are two additional anisotropic types of superconducting orders, an $s$-wave and a $d$-wave. The magnetic phase transitions are here of multiple types and are characterized in Figs.~\ref{fig:PDchiJ12}~\&~\ref{fig:PDchiRest}. The eigenvalues are only calculated up until to a strain value where superconductivity is no longer possible.}
\end{figure*}

\section{Phase diagram: strain and doping}\label{sec:PhaseDiagrams}
In the following sections we discuss the phase diagrams obtained by varying the charge doping and applying an increasing compressive strain. For the most realistic parameter range one obtains Fig.~\ref{fig:PDr}. To explore additional effects from chemical doping and to illustrate the richness of similar compounds, additional phase diagrams are shown in Fig.~\ref{fig:PDHund}. All phase diagrams show strain-induced superconductivity for a broad range of parameters. In addition, three main features can be observed. First, the two types of superconductivity found in earlier works~\cite{Yang2014,Nishiguchi2019}, a pseudospin $j=1/2$ $d$-wave and an orbital $s_{\pm}$-wave, are found. In addition, both types can be found when varying only the doping, for a select set of parameters. Second, at high enough strain an orbital selective pairing which favors one of the in-plane directions can be found. This type of order has a larger component originating from the $xz$ orbital than from $yz$, spontaneously breaking the in-plane symmetry. And third, under compressive strain the susceptibility goes from having the largest peaks in the 
pseudospin states to peaks instead originating from the spin states. This shift affects both the magnetic order and the possible pairing symmetries.

Choosing a realistic regime for the phase diagram, there are two criteria for the chosen interaction parameters. The first criterion is that sufficient doping, either hole or electron, should in accordance to experiment, result in a transition out of the magnetic order. The second criterion is that in undoped Sr$_{2}$IrO$_{4}$ the magnetic order persists up to a strain value of $\epsilon \approx -2\%$~\cite{Seo2019, Haskel2020}. For the most realistic values of the Hund’s and spin-orbit coupling, $J_{\textrm{H}}=0.1U$ and $\lambda=0.6$eV, the first criterion is satisfied for $U \approx 2|t|$, shown in Appendix~\ref{sec:AppA}. The realistic phase diagram is presented for $U=1.1$eV$\approx 3 |t|$. Since the model overestimates the orders this choice only satisfies the second criterion. In Appendix~\ref{sec:AppA} the phase diagram for $U=1.4$eV$\approx 4 |t|$ results in the same phase transitions at higher strain values. Complementary mean field calculations, containing magnetic, superconducting as well as other order parameters, yields a qualitatively similar phase diagram, with differences explained in Appendix~\ref{sec:AppB}.

In Fig.~\ref{fig:PDHund}, we also consider a higher Hund's coupling of $J_{\textrm{H}}=0.25U$, with a lower spin orbit coupling of $\lambda=0.5$eV for the following reason. Hole doping via the substitution of iridium atoms for rhodium or ruthenium atoms could modify the effective interaction parameters and spin-orbit coupling.  Some works estimate the spin orbit coupling of iridium, rhodium and ruthenium as $\lambda^{\textrm{Ir}} \approx 0.45$eV, $\lambda^{\textrm{Rh}}=\lambda^{\textrm{Ru}}\approx 0.19$eV respectively~\cite{Lee2021,Zwartsenberg2020,Brouet2021}. Moreover, ruthenium atoms have a higher Hund's coupling of $J_{\textrm{H}} \approx 0.15U-0.2 U$~\cite{Georges2013}. The full set of phase diagrams are presented in Figs.~\ref{fig:PDr} \& \ref{fig:PDHund}, with identified superconducting and magnetic phases.

\begin{figure*}
\centering  
\includegraphics[width=\textwidth]{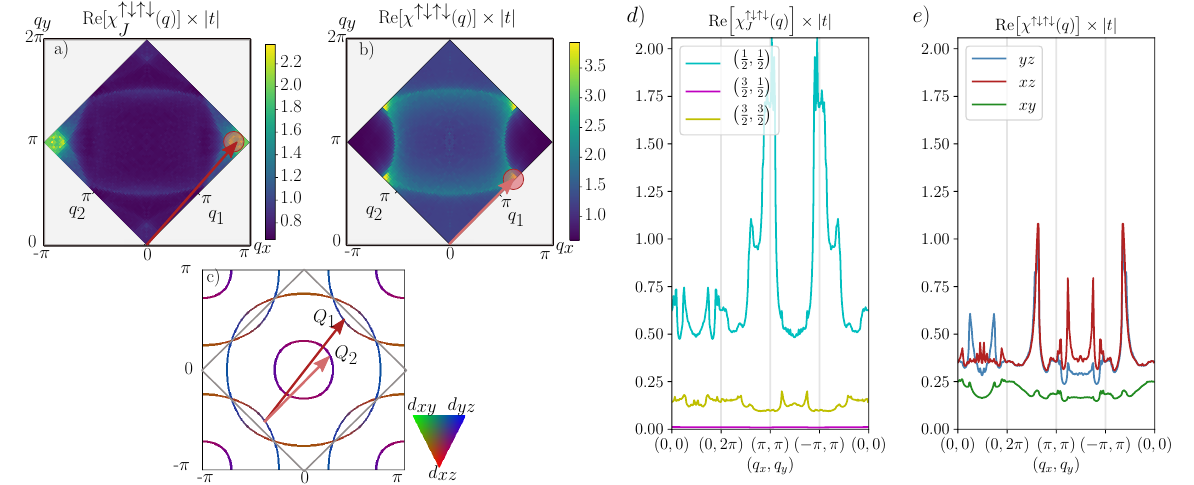}  
\caption{\label{fig:Q1Q2Peaks}The in-plane a) pseudospin $\chi^{\uparrow \downarrow \uparrow \downarrow}_{J}(\boldsymbol{q})$ and b) spin susceptibility $\chi^{\uparrow \downarrow \uparrow \downarrow}(\boldsymbol{q})$, 
Eqs.~\eqref{eq:Chitot}, \eqref{eq:ChiJtot}, for $U=1.1$eV, $J_{\textrm{H}}=0.25U$, $\lambda=0.5$eV, $n=4.8$, $\epsilon=-1\%$. The largest peaks are around $\boldsymbol{Q}_{1} \approx (\pi,\pi)$ and $\boldsymbol{Q}_{2} \approx (\frac{\pi}{2},\frac{\pi}{2})$, respectively. c) Shown for the FS in the extended BZ, $\boldsymbol{Q}_{1}$ connects FS$_{1}$ to itself while $\boldsymbol{Q}_{2}$ connects FS$_{1}$ and FS$_{2}$. On the FS the orbital contributions are given as $|\langle \alpha | \textrm{FS}_{n}| \alpha \rangle|$, for $\alpha= yz, xz, xy$. In d) \& e), the pseudospin susceptibility is split into each $j$-state contribution and the spin susceptibility into that of each orbital. Even though the $j=1/2$ states have the individually largest peaks, the total susceptibility originating from the $yz$- and $xz$-orbitals is larger. Peaks of the type $\boldsymbol{Q}_{1d} \approx (\frac{\pi}{2},\pi)$ are present in the spin susceptibility as peaks belonging entirely to one of the orbitals $yz$ or $xz$.}
\end{figure*}
\section{Magnetic order}\label{sec:MagOrder}
\begin{figure*}
\centering
\includegraphics[width=\textwidth]{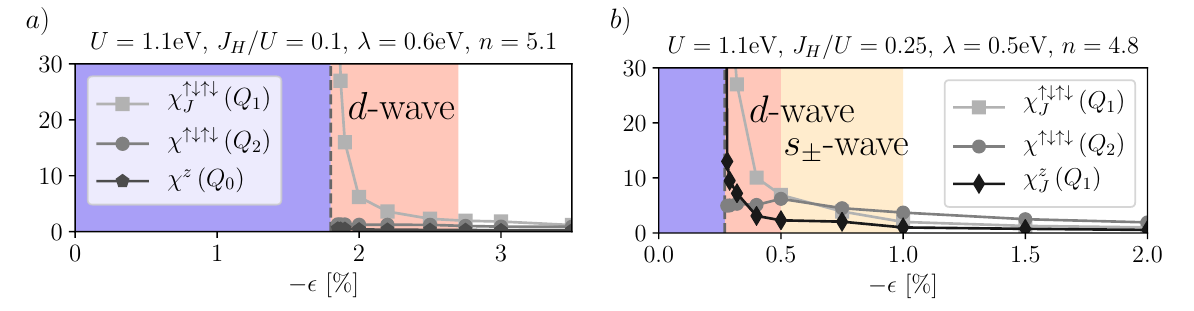}  
\caption{\label{fig:PDchiJ12}Two out of the five types of magnetic instabilities found at the Stoner criterion and the components for the largest susceptibility peaks Re$\left[ \chi \left( \boldsymbol{Q} \right) \right] \times |t|$, are shown along constant doping lines in a) Fig.~\ref{fig:PDr} \& b) Fig.~\ref{fig:PDHund}a. The most prevalent instability occurs in the $j=1/2$ state, in-plane, and with the nesting vector $\boldsymbol{Q}_{1} \approx (\pi,\pi)$: $\chi^{\uparrow \downarrow \uparrow \downarrow}_{J, \left( \frac{1}{2}, \pm \frac{1}{2} \right)} (\boldsymbol{Q}_{1})$. b) At $J_{\textrm{H}} =0.25U$ and hole doping, the $j=1/2$ $\boldsymbol{Q}_{1}$-nesting instability has both in- and out-of-plane components. Close to these instabilities the $\boldsymbol{Q}_{1}$-peaks are the largest. As the strain increases the total in-plane spin susceptibility, Eq.~\eqref{eq:Chitot}, decreases at a slower rate and eventually dominates instead}
\end{figure*}

\begin{figure*}
\centering
\includegraphics[width=\textwidth]{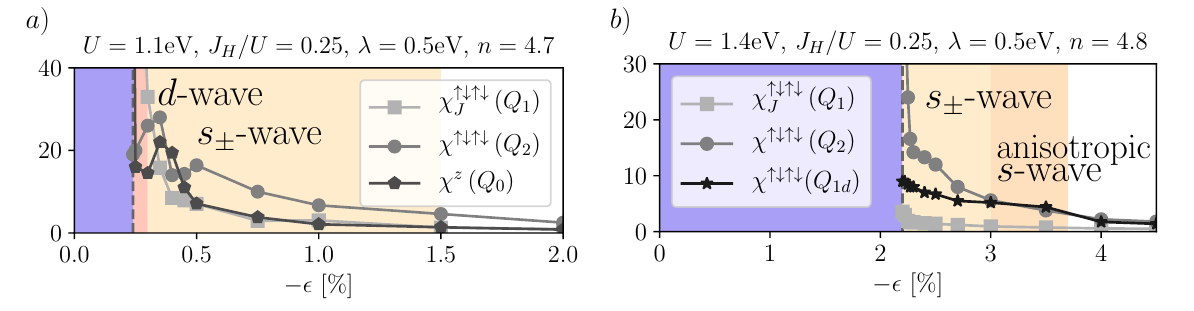}  
\caption{\label{fig:PDchiRest}Additional magnetic instabilities are found for $J_{\textrm{H}} =0.25U$. In a small region around $n=4.7$, in Fig.~\ref{fig:PDHund}a, an out-of-plane ferromagnetic  instability $\chi^{z} (\boldsymbol{Q}_{0})$ accompanies the in-plane order. Note that the competing sizes of different channels here could be an effect of the momentum resolution. In Fig.~\ref{fig:PDHund}b, a spin instability with $\boldsymbol{Q}_{2} \approx (\frac{\pi}{2},\frac{\pi}{2})$, is present for all hole doping. The superconducting $s_{\pm}$-wave is found close to the instability, while the anisotropic $s$-wave appears when the $\boldsymbol{Q}_{1d}$-peak, in the $xz$-orbital, becomes equal in size.}
\end{figure*}

In the RPA calculation, the particle-hole order is found using the Stoner criterion, Eq.~\eqref{eq:StonerDef}. An order can be characterized by two features of $\hat{\chi}(\boldsymbol{q})$. First is the nesting vector $\boldsymbol{Q}=(q_{x}, q_{y})$, given in the extended BZ, at which the Stoner criterion is met. The instabilities in the phase diagrams all appear at the four points, $\boldsymbol{Q}_{1} \approx  (\pi,\pi)$, $\boldsymbol{Q}_{2} \approx (\frac{\pi}{2},\frac{\pi}{2})$, $\boldsymbol{Q}_{0} \approx (0,0)$, and $\boldsymbol{Q}_{1d} \approx (\frac{\pi}{2},\pi)$. The exact location of the instabilities is shifted a small distance from the ideal values, which depends on doping as well as strain. Examples of the dominant peaks are shown in Fig.~\ref{fig:Q1Q2Peaks}. The most relevant components of the susceptibility are shown in Figs.~\ref{fig:PDchiJ12} \& \ref{fig:PDchiRest}, along a few cuts in the phase diagrams. For the lower Hund's coupling in Fig.~\ref{fig:PDr} the only instability is at $\boldsymbol{Q}_{1}$. An order described in real space by a two site unit cell in a square lattice will have a reduced Brillouin zone with a unit vector $\boldsymbol{Q} = (\pi,\pi)$. As antiferromagnetism is a two site order expected in this compound, the $\boldsymbol{Q}_1$ susceptibility peaks are expected to be antiferromagnetic. In Fig.~\ref{fig:PDHund} the $\boldsymbol{Q}_{2}$-instability occurs at most doping values. A real space order corresponding to $\boldsymbol{Q}=(\frac{\pi}{2},\frac{\pi}{2})$, will have a unit cell containing 4 sites. However, it should be noted that the peak $\boldsymbol{Q}_{2}\approx (\frac{\pi}{2},\frac{\pi}{2})$ is doping dependent and never occurs exactly at this value. It is an incommensurate order closer to $(0.6\pi,0.6\pi)$ for hole doping and $(0.4\pi,0.4\pi)$ for electron doping. In Fig.~\ref{fig:Q1Q2Peaks} there is an additional copy of these peaks that connects different copies of pockets in the extended BZ. Another instability, at $\boldsymbol{Q}_{1d}$, only occurs at the highest strains and electron dopings considered. This nesting vector connects segments on the FS with either a clear $xz$-character to other segments belonging to the same orbital. The quasi-1d dispersion of the $xz$ orbital, with $t_{1} \gg t_{\delta}$ in Eq.~\eqref{eq:kinTerms}, leads to the susceptibility in the intra-$xz$ channel having peaks connecting point in momentum space mainly along the $x$-direction. Finally, a ferromagnetic instability at $\boldsymbol{Q}_{0}$ is possible in Fig.~\ref{fig:PDHund}a at $n=4.7$.

The second feature that characterizes a magnetic instability is the channel in which the instability occurs. In Eqs.~\ref{eq:chiz} \& \ref{eq:chiinPlane}, the spin susceptibility $\hat{\chi}(\boldsymbol{q})$, as well as the pseudospin susceptibility $\hat{\chi}_{J}(\boldsymbol{q})$, are divided into magnetic channels. Most instabilities found in Figs.~\ref{fig:PDr} \& \ref{fig:PDHund} are of the type $\chi^{\uparrow \downarrow \uparrow \downarrow}_{J, \left( \frac{1}{2}, \pm \frac{1}{2} \right)} (\boldsymbol{Q}_{1})$, an in-plane magnetic instability with mainly $j=1/2$ contributions. This can be interpreted as the canted in-plane antiferromagnetic order (x-cAFM) observed in each layer experimentally. As denoted in Fig.~\ref{fig:PDchiJ12}b, the higher Hund's coupling, $J_{\textrm{H}}=0.25U$ and $U=1.1$eV, results in an additional out-of-plane component accompanying the in-plane order, with $\chi^{z}_{J, \left( \frac{1}{2}, \pm \frac{1}{2}\right)} (\boldsymbol{Q}_{1})$.

At any point where the $\boldsymbol{Q}_{2}$ instability is present, it occurs in channels of the spin susceptibility, rather than pseudospin. The instability is in-plane and has contributions mainly from the $yz$ and $xz$ orbitals:  $\chi^{\uparrow \downarrow \uparrow \downarrow}_{yz} (\boldsymbol{Q}_{2})$ \& $\chi^{\uparrow \downarrow \uparrow \downarrow}_{xz} (\boldsymbol{Q}_{2})$. Even though this instability is only present at $U=1.4$eV and $J_{\textrm{H}}=0.25U$, these susceptibility peaks remain large in the entire Fig.~\ref{fig:PDHund}a phase diagram. At high hole doping in Fig.~\ref{fig:PDHund}a a purely out-of-plane ferromagnetic instability $\chi^{z} (\boldsymbol{Q}_{0})$ occurs for spins in each of the orbitals $yz$, $xz$, and $xy$. 

As the strain increases the peaks in the susceptibility at different channels decrease at different rates. Moreover, we will see below that superconductivity found directly adjacent to a magnetic order is mediated by those magnetic fluctuations while superconducting orders which are mediated by other fluctuations can become more favorable as strain is increased further. In Fig.~\ref{fig:PDchiJ12}b, at $J_{\textrm{H}}=0.25U$, the spin susceptibility peaks decrease significantly slower than those of the antiferromagnetic pseudospin order. At strain $\epsilon= -0.5\%$, and beyond, the peaks in spin susceptibility become dominant. The effective particle-particle vertex, Eq.~\eqref{eq:VppEff}, develops peaks at the same $\boldsymbol{Q}$-points as for the spin (or pseudospin) susceptibility. To track which magnetic fluctuation mediates the superconducting orders, the strengths of spin and pseudospin fluctuations are compared. The total susceptibilities in the two bases are
\begin{equation}\label{eq:Chitot}
\chi^{\sigma \sigma' \sigma \sigma'}(\boldsymbol{q}) = \frac{1}{2} \sum_{s,s'} \sum_{\alpha, \beta} e^{i q_{x} (\Theta_{s} - \Theta_{s'} )}  \chi^{(\alpha \sigma) (\alpha \sigma') (\beta \sigma) (\beta \sigma')}_{sss's'}(\boldsymbol{q})
\end{equation}
\begin{equation}\label{eq:ChiJtot}
\chi^{\tau \tau' \tau \tau'}_{J}(\boldsymbol{q}) = \frac{1}{2} \sum_{s,s'} \sum_{m,n} e^{i q_{x} (\Theta_{s} - \Theta_{s'} )}  \chi^{(m \tau) (m \tau') (n \tau) (n \tau')}_{J,sss's'}(\boldsymbol{q})
\end{equation}
where $\Theta_s$, with $s=A,B$, chooses the sublattice such that $\Theta_{A} =0$ and $\Theta_{B} =1$.

It should be noted, that the pseudospin susceptibility nesting vector $\boldsymbol{Q}_{1}$ can be related to the hidden spin density wave (hSDW) orders found in other models for multi-band superconductors~\cite{Berg2012,Rodriguez2020}. The $j$-state basis mixes spin and orbital degrees of freedom and therefore the peaks found correspond to a linear combination of channels that favors both SDW and hSDW orders.

\section{Superconductivity}\label{sec:Supercond}
\begin{table*}
\centering   
\begin{tabular}{|l|c|c|c|c|}
\hline
Symmetry $\eta^{\mu}_{R} (\boldsymbol{k})$ & $R=0$ & $R=1$                          & $R=2$                                             & $R=3$                          \\ \hline
A$_{1g}$, $s$                              & 1     & $\cos k_{x} + \cos k_{y} $     & $2 \cos k_{x} \cos k_{y}$                           & $\cos 2 k_{x} + \cos 2 k_{y} $ \\ \hline
B$_{1g}$, $d_{x^{2}- y^{2}}$               &   -    & $\cos k_{x} - \cos k_{y}$      &       -                                            & $\cos 2 k_{x} - \cos 2 k_{y} $ \\ \hline
B$_{2g}$, $d_{xy}$                         &   -    &             -                   & $2 \sin k_{x} \sin k_{y} $                          &               -                 \\ \hline
E$_{u}$, $p$                                     &     -  & $ \sin k_{x}, \sin k_{y} $ & $\sin ( k_{x} +k_{y}), $ & $\sin 2 k_{x}, \sin 2 k_{y} $  \\ 
& & & $\sin ( k_{x} - k_{y}) $ & \\ \hline
\end{tabular}
\caption{\label{tab:SymIrr} The lattice harmonics for each relevant symmetry on the square lattice is given for different radii $R$, describing what distance neighbors the symmetry is found on. $R=0$ is on-site, $R=1$ is nearest neighbors, and so on.}
\end{table*}
\subsection{Symmetries}\label{sec:Symmetries}
In single-orbital models, it is of highest importance to determine whether the superconductivity is a spin-singlet or a spin-triplet order. Topological superconductivity and Majorana modes arise from superconductivity with $p$-wave pairing, which requires spin-triplet pairing in those systems. Efforts to induce superconductivity through the proximity effect are thus often focused on finding spin-triplet orders. However, once multiple orbitals and spin-orbit coupling are considered the connection between spin-triplets and $p$-wave symmetry is no longer a strict requirement. Multi-orbital models allow for a large set of possible pairing symmetries. The pairing matrix $\hat{\Delta}(\boldsymbol{k})$ must be antisymmetric under the full $\mathcal{S} \mathcal{P} \mathcal{O} \mathcal{T}$-exchange~\cite{Sigrist1991,Linder2019}. Therefore, any pairing can be classified as being either even or odd under spin exchange ($\mathcal{S}$), relative coordinate reflection ($\mathcal{P}$), orbital exchange ($\mathcal{O}$), and relative time exchange ($\mathcal{T}$) as defined in Appendix~\ref{sec:AppD}. As only the static case is considered in this work the order parameter is constant, and therefore even, under $\mathcal{T}$. Note that the operators $\mathcal{P}$ and $\mathcal{T}$ only exchange relative parameters, and are thus different from the reflection and time reversal operators.

For example, classifying symmetries for strong spin-orbit coupling in the predicted $j=1/2$ $d$-wave, will inherently result in both spin-singlet and spin-triplet contributions. The pairing is more accurately described by the total angular momentum of the pair, which can take the values $J=0,1,2,3$~\cite{Dutta2021,Adarsh2022}. Only within the $j_{1} \otimes j_{2} =\frac{1}{2} \otimes \frac{1}{2}$ sector do we still only get pairs that are either a $J=0$ singlet or triplets $J=1$ with $M=-1,0,+1$. In Appendix~\ref{sec:AppD} the symmetry operations in the orbital basis are shown for the two types of pairing found in Fig.~\ref{fig:PDr}.

\begin{figure*}
\centering    
\includegraphics[width=\textwidth]{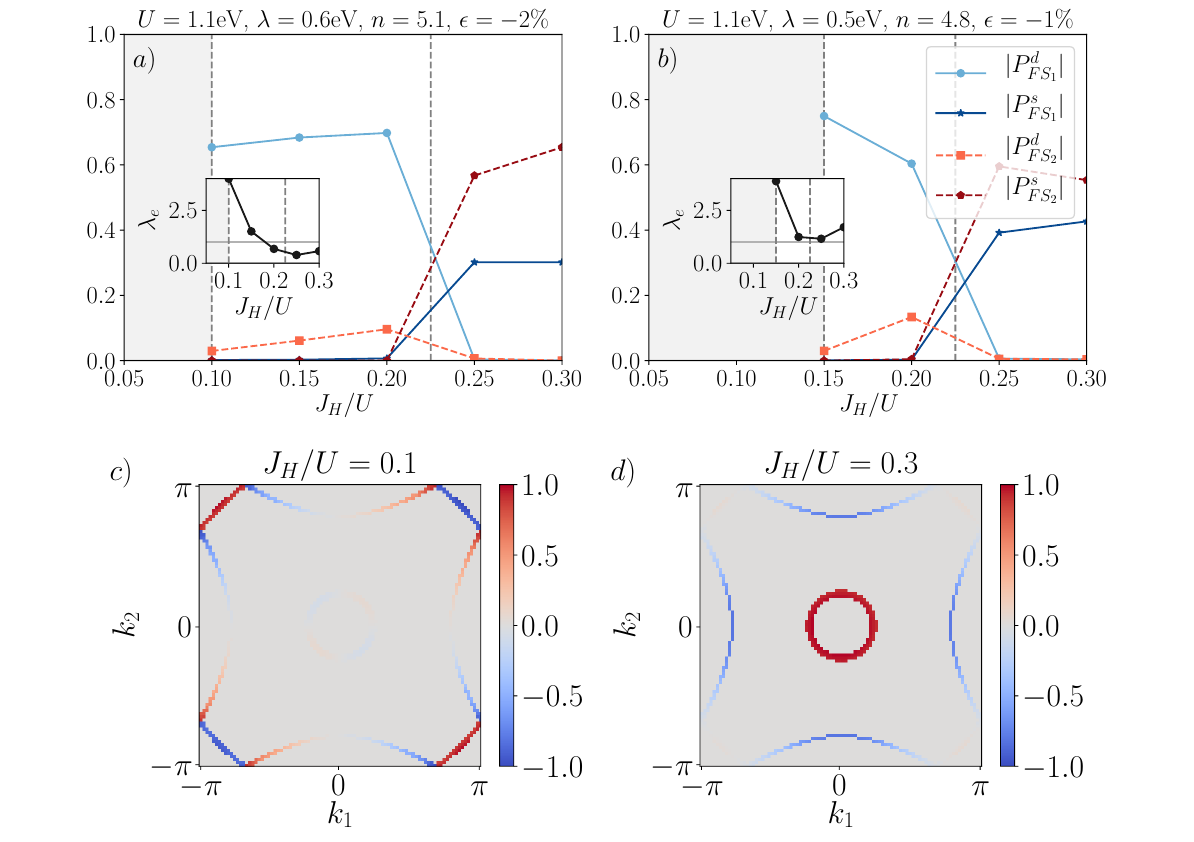} 
\caption{\label{fig:JHvarEigSymU11}Both the dominant symmetry and the relative weight on the pockets change as the Hund's coupling is varied. In a) and b) the spin-orbit coupling $\lambda$ is fixed and the spin-singlet order for each pocket at the Fermi surface is projected onto each spatial symmetry, as in Eq.~\eqref{eq:symProj}. The inserts show the largest eigenvalue $\lambda_{e}$ of the linearized gap equation Eq.~\eqref{eq:SCeq}. c) and d) show $\Delta^{s/t}_{\textrm{FS}_{n}}(\boldsymbol{k})$ for two values of $J_{\text{H}}$ for the FS belonging to a). In the $d$-wave state, the weight on the FS$_{1}$ is largest. For the $s$-wave state, the opposite is true and pockets have opposite signs, identifying it as an $s_{\pm}$-wave.}
\end{figure*}

The spatial symmetry can be considered for the non-interacting bands $n_{\tau}$ by projecting the intraband pairing onto the Fermi surface (FS). The FS may contain three types of pockets belonging to two types of bands. The larger pockets, consisting mainly of $(j,j_{z})=(\frac{1}{2}, \pm \frac{1}{2})$ states, are located around the points $(k_{1}, k_{2})=(\pi, \pi),  (\pi,0)$ and have superconducting gaps that can be described by the same spatial symmetry. We therefore only look at one of these pockets, denoted FS$_{1}$. The smaller FS$_{2}$, with mainly $(j,j_{z})=(\frac{3}{2}, \pm \frac{3}{2})$ is centered around $(k_{1}, k_{2})=(0,0)$. The spatial symmetry is thus studied for four intraband parameters: pseudospin-singlets $\Delta^{s}_{\textrm{FS}_{1}}(\boldsymbol{k})$, $\Delta^{s}_{\textrm{FS}_{2}}(\boldsymbol{k})$ and pseudospin-triplets $\Delta^{t}_{\textrm{FS}_{1}}(\boldsymbol{k})$, $\Delta^{t}_{\textrm{FS}_{2}}(\boldsymbol{k})$. $\Delta^{s/t}_{\textrm{FS}_{n}}(\boldsymbol{k})$ is the order parameter Eq.~\eqref{eq:SCeq} projected onto the band at the Fermi surface $FS_n$. The number of points belonging to a pocket is $N_{\textrm{FS}_{n}}$. For example the pseudospin-singlet is
\begin{equation}
\begin{aligned}
    \Delta^{s}_{\textrm{FS}_{n}}(\boldsymbol{k})  = &\sum_{\boldsymbol{k} \in \textrm{FS}_{n}} \sum_{\alpha \beta} \sum_{m m'}  \left[ U_{\boldsymbol{k}}^{\dagger} \right]_{n \alpha } \left[ U_{-\boldsymbol{k}}^{\dagger} \right]_{n \beta } \\ &\times \frac{1}{\sqrt{2}}  \left( M_{\alpha (m +)}^{\dagger} M_{\beta (m' -)}^{\dagger}  \Delta_{(m +) (m' -)} (\boldsymbol{k}) \right. \\ & \left. - M_{\alpha (m -)}^{\dagger} M_{\beta (m' +)}^{\dagger} \Delta_{(m-) (m' +)} (\boldsymbol{k}) \right)
\end{aligned} 
\end{equation}
where $M$ is the matrix in Eq.~\eqref{eq:Mmat} and the matrix $U_{\boldsymbol{k}, \alpha n}$ transforms the band basis into the orbital basis. In the numerical calculations, there are small but non-zero interband contributions that will not be considered further. The spatial symmetries can be quantified via the projection coefficients $P^{\mu,R}_{s/t, \textrm{FS}_{n}}$ onto the basis functions for each parity irreducible representation, $\eta^{\mu}_{R}(\boldsymbol{k})$, as given in Table~\ref{tab:SymIrr}
\begin{equation}
\Delta^{s/t}_{\textrm{FS}_{n}}(\boldsymbol{k})= \sum_{\mu, R} P^{\mu,R}_{s/t, \textrm{FS}_{n}} \eta^{\mu}_{R}(\boldsymbol{k}).
\end{equation}
The projection coefficient for each irreducible representation is thus found via
\begin{equation}\label{eq:symProj}
P^{\mu,R}_{s/t, \textrm{FS}_{n}}= \frac{1}{N_{\textrm{FS}_{n}}} \sum_{\boldsymbol{k} \in \textrm{FS}_{n}} \eta^{\mu}_{R}(\boldsymbol{k}) \Delta^{s/t}_{\textrm{FS}_{n}}(\boldsymbol{k}).
\end{equation}
Examples of the two leading types of pairings are shown in Fig.~\ref{fig:JHvarEigSymU11}. The most prominent $d$-wave is a $j=1/2$ pseudospin singlet with a $\eta^{B_{1g}}_{R=1}$ pairing. The found $s_{\pm}$-wave order has contributions from both $\eta^{A_{1g}}_{R=0} (\boldsymbol{k})$ and $ \eta^{A_{1g}}_{R=2} (\boldsymbol{k}) $, with opposite sign for the two pockets. The main $j$-states components are from $(\frac{1}{2}, \pm \frac{1}{2})$ and $(\frac{3}{2}, \pm \frac{3}{2})$. However, components mixing $j$-states is stronger than for the $d$-wave. The symmetry in terms of orbital origin is specified further in Appendix~\ref{sec:AppD}.

\subsection{Realistic strain-induced order}
At low Hund's coupling and high SOC, a $d$-wave is found for a wide range of doping values once there is no longer a magnetic order present. The mean field calculations in Appendix~\ref{sec:AppB} corroborate the prediction of this order. The magnetic region extends up to $\epsilon=-1.5\%$ at the undoped $n=5$, while it persists at higher strains on the electron doped side. Because of the required compressive strain, two bands are present at the FS at all points where superconductivity is found. As seen in Fig.~\ref{fig:JHvarEigSymU11}, the gap on FS$_{1}$ is significantly larger than on FS$_{2}$. The $d$-wave originates from the $j=1/2$ states, which are the states the band at FS$_{1}$ also belongs to. Strain has increased the size of the $j=1/2$ electron pocket, FS$_{1}$, in the entire region where superconductivity is found. Even for hole doping at $\epsilon=-2\%$ the pocket FS$_{1}$ is comparable in size to FS$_{1}$ at $\epsilon=0$ and electron doping, as further explained in Appendix~\ref{sec:AppB}.

\begin{figure}
\centering   
\includegraphics[width=\linewidth]{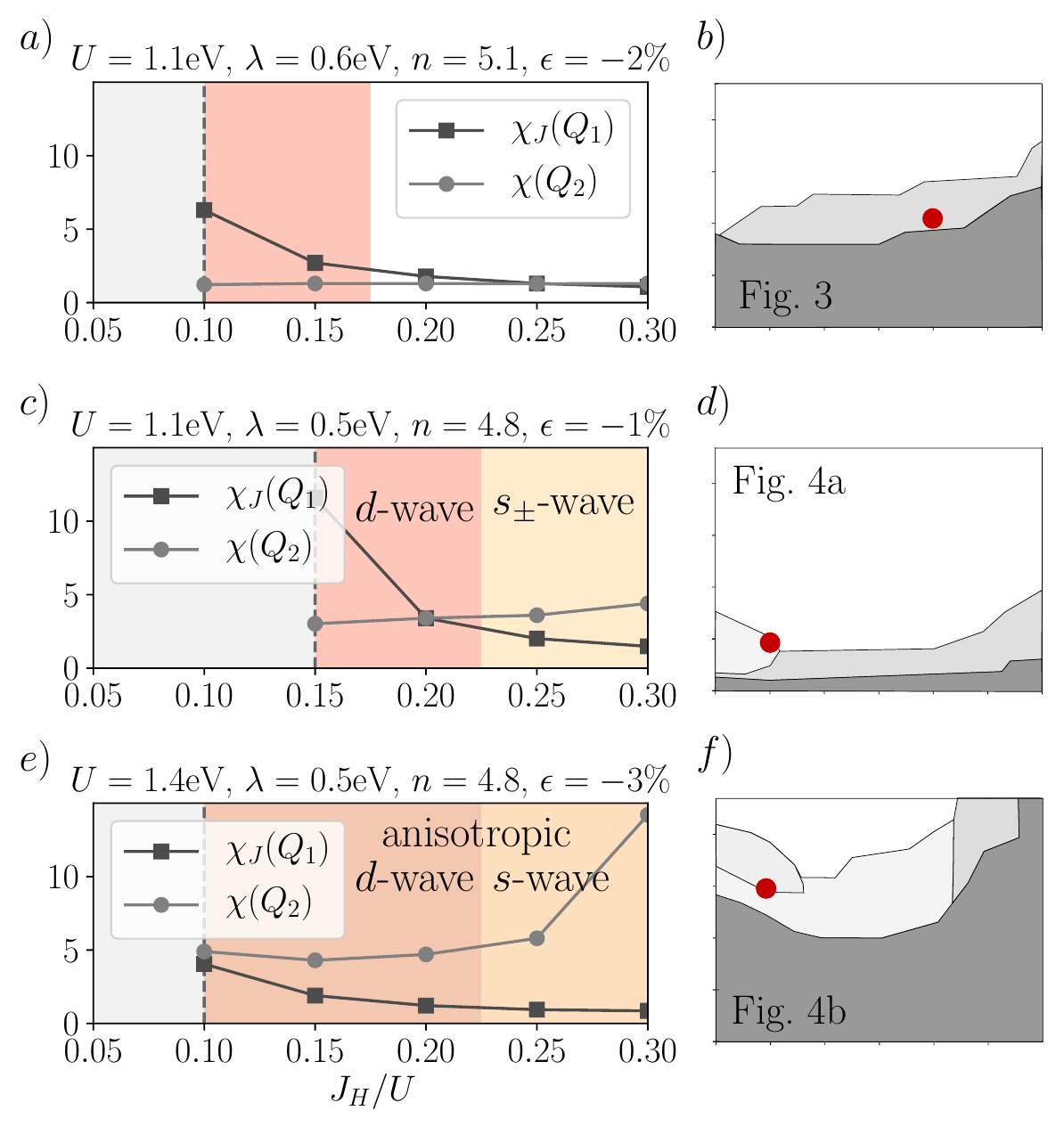}  
\caption{\label{fig:maxPeaks}Maximum peaks in the spin and pseudospin susceptibility, Eqs.~\eqref{eq:Chitot} \& \eqref{eq:ChiJtot}, for the same calculation as in Fig.~\ref{fig:JHvarEigSymU11} and for $U=1.4$eV. Each value is the in-plane Re$\left[ \chi^{\uparrow \downarrow \uparrow \downarrow} \left( \boldsymbol{Q} \right) \right] \times |t|$. For a Hund's coupling where a pseudospin $d$-wave is favored $\chi_{J} (\boldsymbol{Q}_{1}) > \chi (\boldsymbol{Q}_{2})$. The location in the phase diagrams for the points chosen in each plot is shown in the lower row.}
\end{figure}
\subsection{Varying Hund's coupling}\label{sec:Hunds}
There are several important factors that determine which pairing symmetry is favored. Doping and strain change the shape of the Fermi surface, and thus the nesting vectors $\boldsymbol{Q}$, as well as the orbital contributions in each pocket. Both compressive strain and hole doping increase the presence of $yz$ and $xz$ orbitals. However, the type of fluctuations which dominate the RPA interaction vertex depends on the interaction parameters. In Fig.~\ref{fig:JHvarEigSymU11} the largest eigenvalue of Eq.~\eqref{eq:SCeq}, and the symmetry of the pairing, are shown as the Hund's coupling varies. We observe a general trend in which the $s_{\pm}$-order becomes more favorable than $d$-wave superconductivity at $J_{\textrm{H}} \geq 0.25U$. For the lower SOC, $\lambda=0.5$eV, the largest eigenvalue is above unity for all values of Hund's coupling considered. The $s_{\pm}$-wave is only possible for low SOC and hole doping, since that places the Fermi level deeper in the band of $(j, j_{z})=(\frac{3}{2}, \pm \frac{3}{2})$ character. By contrast, the $d$-wave order is present for a wider range of parameters.

At all points there are two main types of competing fluctuations; pseudospin fluctuations around $\boldsymbol{Q}_{1} \approx (\pi,\pi)$ and spin fluctuations around $\boldsymbol{Q}_{2} \approx (\frac{\pi}{2},\frac{\pi}{2})$. By studying the maximum peak values in Fig.~\ref{fig:maxPeaks} one can determine which one of these fluctuations best describes the system. As can be seen in Figs.~\ref{fig:PDchiJ12} \& \ref{fig:PDchiRest}, dominating spin fluctuations promote an $s_{\pm}$-wave order. However, if both types of fluctuations are of roughly equal size the $j=1/2$ $d$-wave is favored.

\subsection{Multi-band orders}
\begin{figure}
\centering    
\includegraphics[width=\linewidth]{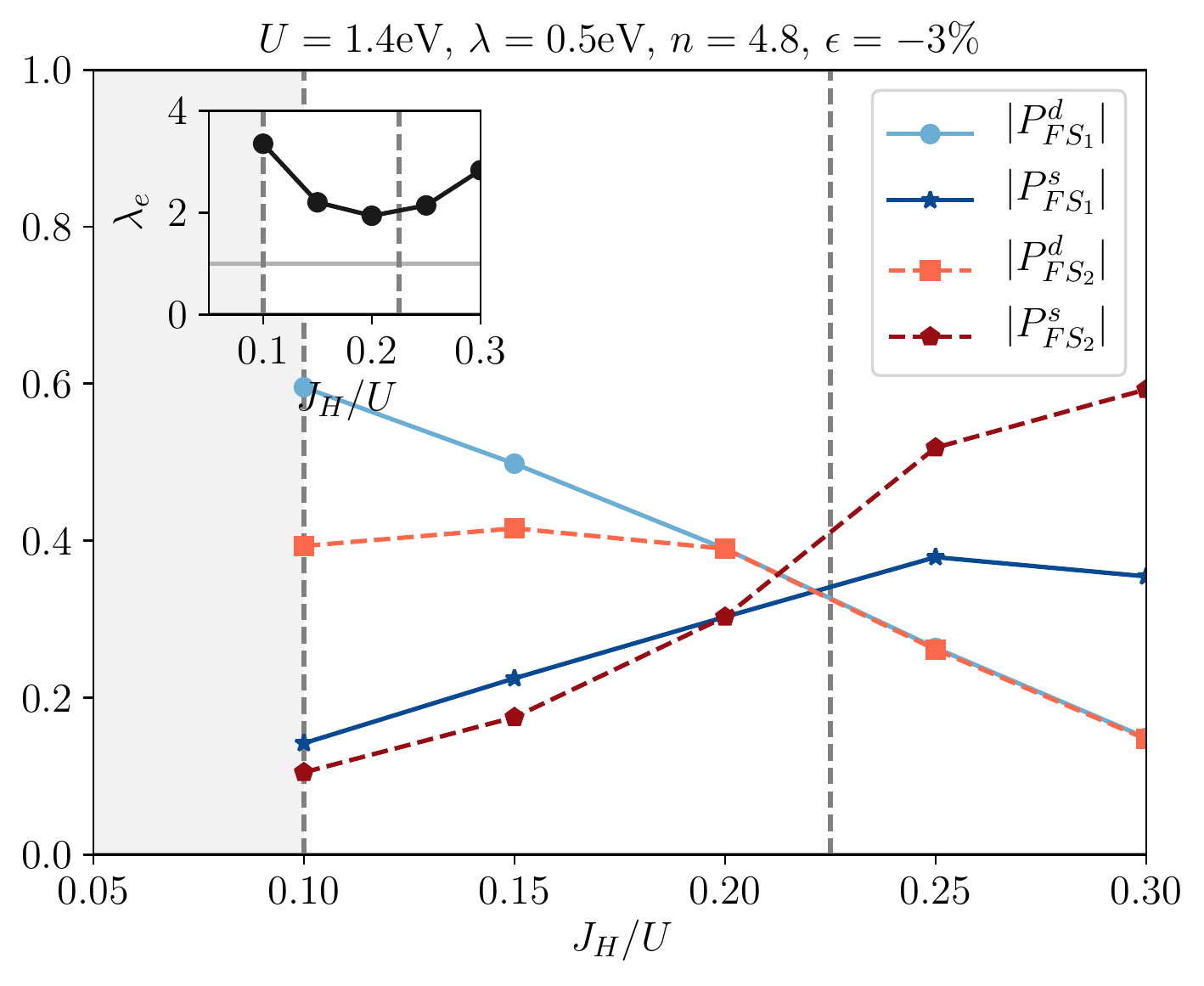} 
\caption{\label{fig:JHvarEigSymla05}For a higher $U=1.4$eV and high compressive strain $\epsilon=-3\%$ two new anisotropic orders are found. In Appendix~\ref{sec:AppE} the origin is identified as a higher contribution to the pairing from one of the orbitals ($xz$). For the pairing at the Fermi surface this manifests as both orders being a mix of $s$- and $d$-wave symmetries. As the Hund's coupling is increased the order goes from being predominantly a $d$-wave order, with nodes, to a node-less order with stronger $s$-wave components. The insert shows the largest eigenvalue $\lambda_{e}$.}
\end{figure}
For a higher Hund's coupling superconducting orders are favored which open large gaps on multiple pockets at the Fermi surface. In Fig.~\ref{fig:PDHund}a, there are two distinct superconducting orders at $U=1.1$eV. The magnetic order disappears for small strains and superconductivity is only possible up to $\epsilon= -2\%$. Moreover, the symmetry of the superconducting order is dependent on doping. At high hole doping and some strain the pairing is an $s_{\pm}$-wave, while remaining a $d$-wave for all other doping values. When we set the Hubbard $U$ to a higher value of $U \approx 4|t|$, as seen in Fig.~\ref{fig:PDHund}, and turn on a high compressive strain the band structure changes.  This can lead to new pairing functions which were not seen earlier. A drastic change occurs in Fig.~\ref{fig:PDHund}b where some regions have an anisotropic $s$- or $d$-wave pairing. For all values of $J_{\textrm{H}}$ considered in Fig.~\ref{fig:JHvarEigSymla05}, the anisotropic order, which is a mix of $s$- and $d$- wave pairing, is found. The $s$-wave contribution increases with Hund's. The orbital components, as well as a simple model for this state, are described in Appendix~\ref{sec:AppE}. This superconducting order is an orbital-selective state with a stronger spin-singlet in the $xz$-orbital. One notable reason for this new type of pairing is the increased spin nature of the fluctuations as compressive strain is increased. As already discussed, all regions with anisotropic pairing are mediated by a large spin susceptibility peak in $\chi^{\uparrow \downarrow \uparrow \downarrow}_{xz}(\boldsymbol{Q}_{1d})$. In Fig.~\ref{fig:maxPeaks}, the larger spin to pseudospin susceptibility ratio can be compared for the increased strain, at all $J_{\textrm{H}}$ values.

\subsection{Odd parity}\label{sec:OddParity}
\begin{figure}
\centering   
\includegraphics[width=\linewidth]{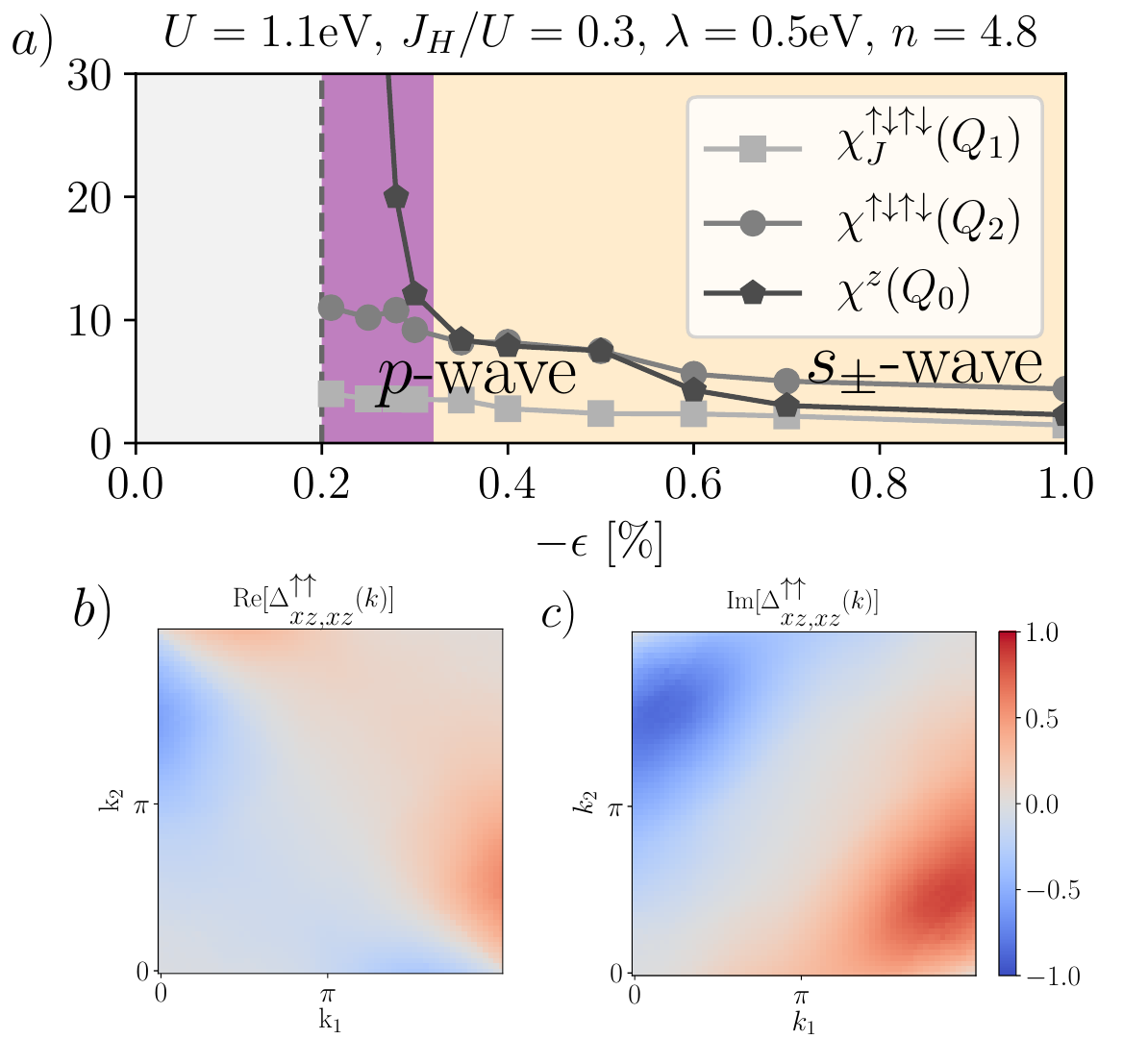}
\caption{\label{fig:maxPeakspwave}a) The total in- and out-of-plane susceptibility, Eqs.~\eqref{eq:Chitot} \& \eqref{eq:ChiJtot}, are shown for a higher $J_{\textrm{H}}=0.3U$, where $\chi^{z}(\boldsymbol{Q}_{0})$ is the largest component. Only for a small region close to the magnetic instability is an odd parity $p$-wave pairing favored. b) \& c) $\Delta^{\uparrow \uparrow}_{xz,xz}(\boldsymbol{k})$ of the $p$-wave pairing is shown, revealing a dominating $p + ip$ structure. The equally large orbital component is $\Delta^{\uparrow \uparrow}_{yz,yz}(\boldsymbol{k})= -\Delta^{\uparrow \uparrow}_{xz,xz}(\boldsymbol{k})$.}
\end{figure}

In general, the out-of-plane $\chi^{z}(\boldsymbol{Q}_{0})$ susceptibility peaks remain smaller than either the in-plane spin $\chi^{\uparrow \downarrow \uparrow \downarrow} \left( \boldsymbol{Q} \right)$ or pseudospin $\chi^{\uparrow \downarrow \uparrow \downarrow}_{J} \left( \boldsymbol{Q} \right)$ peaks. This is the case for all values calculated so far, except for the one small region in Fig.~\ref{fig:PDchiRest} with large hole doping, high Hund's coupling, and low compressive strain. A high enough Hubbard interaction $U \geq 1.1$eV is also required. As several of these factors also increase the in-plane spin susceptibility, the out-of-plane susceptibility only dominates in a very small parameter range. In Fig.~\ref{fig:maxPeakspwave} one such small patch can be found for a very high Hund's coupling, $J_{\textrm{H}}=0.3U$. Accompanying these fluctuations is an odd parity $p$-wave superconductivity. The main contribution, described by the symmetry representation detailed in Appendix~\ref{sec:AppD}, is
\begin{equation}
    \Delta_{p}(\boldsymbol{k}) \approx \left( h_{x} \otimes \sigma^{z} + i h_{y}  \otimes \mathbb{I} \right)  \otimes \lambda_{3}.
\end{equation}
where $\lambda_j$ is the $j^{th}$ Gell-Mann matrix~\cite{GellMann1961} acting in the three-orbitals space.
The $p$-wave pairing has several leading terms, which are $h_{x} =  -\sin k_{x}+ \sin k_{x} \cos k_{y} + \sin(2 k_{x})$ and $h_{y} =  -\sin k_{y}+ \sin k_{y} \cos k_{x} + \sin(2 k_{y})$. Some smaller terms are proportional to $\left( h_{x} \otimes \sigma^{z} + i h_{y}  \otimes \mathbb{I} \right)  \otimes \lambda_{1} $ and $\Delta^{\downarrow \downarrow}_{xy,xy}(\boldsymbol{k}) \propto \Delta^{\uparrow \uparrow}_{yz,yz}(\boldsymbol{k})$. Up- and down-spin pairing have the opposite chirality. This helical $p$-wave thus preserves time-reversal symmetry (TRS), since $\Delta_{\alpha \beta}^{\downarrow \downarrow} (\boldsymbol{k}) = \left(\Delta_{\alpha \beta}^{\uparrow \uparrow} \right) ^{\ast}(-\boldsymbol{k})$. The helical nature is preserved for both Fermi surfaces, FS$_1$ and FS$_2$, with a larger gap on FS$_{2}$. A $\mathbb{Z}_{2}$ topological invariant can therefore be determined. We can consider a Chern number for each pseudospin sector, as outlined in Appendix~\ref{sec:AppF}. The defining features of the pairing are
\begin{itemize}
    \item The helical $p$-wave preserves time reversal symmetry, such that the total Chern number vanishes, $\mathcal{C}_{\textrm{tot}}= 0$. 
    \item The pseudospin Chern number, $\nu = (\mathcal{C}_{+} - \mathcal{C}_{-})/2$, also vanishes such that in each pseudospin sector $\mathcal{C}_{\tau} =0$.
    \item In each pseudospin sector we define the pocket Chern number $\mathcal{C}_{n,\tau}$ for the band $n$. We find $\mathcal{C}_{n,\tau} = \pm 1$ such that for each band the pseudospin Chern number is $\nu_{n} = \pm 1$. The pocket contributions cancel such that $\nu = \sum_{n} \nu_{n} =0$, as the pocket of $j=3/2$- character has the opposite chirality to the $j=1/2$ pocket.
    \item Therefore no topologically protected edge/vortex modes are expected.
\end{itemize}
Even for chiral TRS-breaking superconductors the Fermi surface topology can result in a topologically trivial state, when multiple pockets are present~\cite{Sato2009,Farrell2013}. The possible Chern numbers in multiband superconductors depend on the pairing function, and the location in the Brillouin zone of the resulting topological charges~\cite{Raghu2010,Imai2012,Scaffidi2015}.

It might be possible to expand the $p$-wave regime by tuning parameters such that ferromagnetism is favored. In our model, increasing the Hubbard coupling, $U$, accomplishes this. However, with a higher $U$, a higher strain required to reach the superconducting regime but a higher $U$ also increases the in-plane fluctuations, as in Fig.~\ref{fig:PDchiRest}. This is caused by the other fluctuations being favored when the pocket with $j=3/2$ states becomes large. Only for a smaller pocket and a large $U$ would the $p$-wave be favored. The odd parity order is thus a fine-tuned case which is found beyond realistic parameters.

\section{Discussion}\label{sec:Discussion}
While strong interactions, spin-orbit coupling, and the proximity of multiple $d$-bands to the Fermi level all point to the possibility of unconventional superconductivity, experimental observation of superconductivity is still missing.  In this work we suggest that compressive strain may be a possible knob that, together with doping, can turn the system from magnetic to superconducting.  We model the system using the extended Hubbard-Kanamori Hamiltonian and map out its phase diagram. Magnetic orders are found using the Stoner criterion while superconductivity is studied using the RPA linearized Eliashberg equation. For the range of parameters considered in this work, we find prominent regions of strain-induced superconductivity, among them a large fraction exhibits $d$-wave pairing.

The $d$- and $s_{\pm}$-wave orders found are the same as in previous studies of the unstrained compound. For the values considered here the $s_{\pm}$-wave can arise adjacent to the AFM order, for a high enough Hund's coupling. For high strains and Hund's coupling, new orbital-selective, anisotropic $s$- or $d$-wave orders are found. In addition, we find an out-of-plane ferromagnetic order. In a very fine-tuned region, the out-of-plane susceptibility mediates an odd parity $p$-wave order. We note that the work of Ref.~\cite{Meng2014} finds a $p$-wave order in hole doped Sr$_{2}$IrO$_{4}$, albeit at an extremely large Hubbard coupling, $U=12t$, and a lower Hund's coupling of $J_{\textrm{H}}=0.15U$. The odd parity order we find is favored by a high Hubbard coupling, $U$, in the unstrained compound and could be the dominant order at those values. At higher strains the ferromagnetic fluctuations vanish. The $p$-wave order is found to be helical and topologically trivial, as determined via the $\mathbb{Z}_{2}$ invariant. However, other values of the Hund's coupling and different Fermi surface geometry could potentially break time reversal symmetry and result in a chiral $p$-wave~\cite{Scaffidi2014}. As in the case of Sr$_{2}$RuO$_{4}$, the topological nature of such a state in the iridates is highly dependent on the Fermi surface geometry and orbital composition.

It should be mentioned that the type of possible magnetic instabilities is not altered by the compressive strain for the realistic value of the Hund's coupling, $J_{\textrm{H}}=0.1U$. However, with increasing compression the pseudospin $j=1/2$ susceptibility decreases faster than the spin susceptibility, bringing the two leading fluctuation peaks to comparable sizes. At high Hund's coupling and lower SOC we find an antiferromagnetic order that can have a mixed in- and out-of-plane structure. Moreover, at higher strains a spin-like incommensurate magnetic order is possible. Further calculations of the particle-hole self-energy are required to characterize this magnetic order. The competition between two types of susceptibility peaks mediating the d- and s-wave orders, shares many similarities with work done on iron-pnictide superconductors~\cite{Graser2009,Maier2011}. In iron pnictides a different nesting vector $\boldsymbol{Q}=(\pi,0)$ connects pockets and mediates the $s_\pm$-wave. Ref.~\cite{Graser2009} predicts nearly degenerate multi-band $s$- and $d$-wave orders where a small change in the interaction parameters determines the favorable order. However, in contrast to our work their model does not contain spin-orbit coupling. In our work, the strong SOC and the consequent pseudospin degrees of freedom are partially responsible for the dominance of the $d$-wave order in the realistic regime phase diagram.

The strain-induced superconducting regimes we find all occur when the Fermi surface has multiple pockets. All superconducting orders are thus multigap orders~\cite{Kresin2013,Zhitomirsky2001}. However, for a higher SOC or lower $U$ that would not necessarily be the case. The size of the second pocket FS$_{2}$ and the value of the Hund’s coupling determine the relative sizes of the gap for the two pockets. In Fig.~\ref{fig:JHvarEigSymU11}, the relative gaps are shown projected onto the FS at low Hund's coupling, and we find that pocket 1 dominates: Max$\left[ \Delta_{\textrm{FS}_{1}}(\boldsymbol{k}) \right] \gg$ Max$\left[ \Delta_{\textrm{FS}_{2}}(\boldsymbol{k}) \right]$. The smaller gap is expected to have a smaller effect on the (shared) critical temperature. Methods to determine the relative size and pairing structures of two gap superconductors have been explored in multiple compounds such as MgB$_{2}$~\cite{Souma2003} and SrTiO$_{3}$~\cite{Fernandes2013}. Further proposals have been made to detect any offset phases between pairing functions. Especially for the $s_{\pm}$-wave order where signatures of the phase difference between the two pockets could be detected via Josephson tunneling~\cite{Chen2010,Berg2011,Rodriguez-Mota2016}.

Our results indicate that the doped and strained regime is of interest for potential iridate superconductivity. Understanding changes to magnetic fluctuations for any experiment combining strain and doping would provide great insight to the interplay of interactions and spin-orbit coupling in transition metal oxides. An observation of the susceptibility peaks that we identify as mediating superconductivity could hint at a possible superconducting phase nearby. Further understanding the signatures of the possible superconductivity, such as the Knight shift in the magnetic susceptibility~\cite{Romer2019} could be a direction for future work.

\section{Acknowledgments}
We acknowledge financial support from NSERC, RQMP, FRQNT, an Alexander McFee Fellowship from McGill University (LE), a Team Research Project from FRQNT, a grant from Fondation Courtois, and a Canada Research Chair. Computations were made on the supercomputers managed by Calcul Qu\'ebec and Compute Canada. The operation of these supercomputers is funded by the Canada Foundation for Innovation (CFI), the ministère de l'\'economie, de la science et de l'innovation du Qu\'ebec (MESI) and the Fonds de recherche du Qu\'ebec - Nature et technologies (FRQ-NT).

\appendix 
\section{Phase diagrams at other U}\label{sec:AppA}
\begin{figure*}
\centering  
\includegraphics[width=\textwidth]{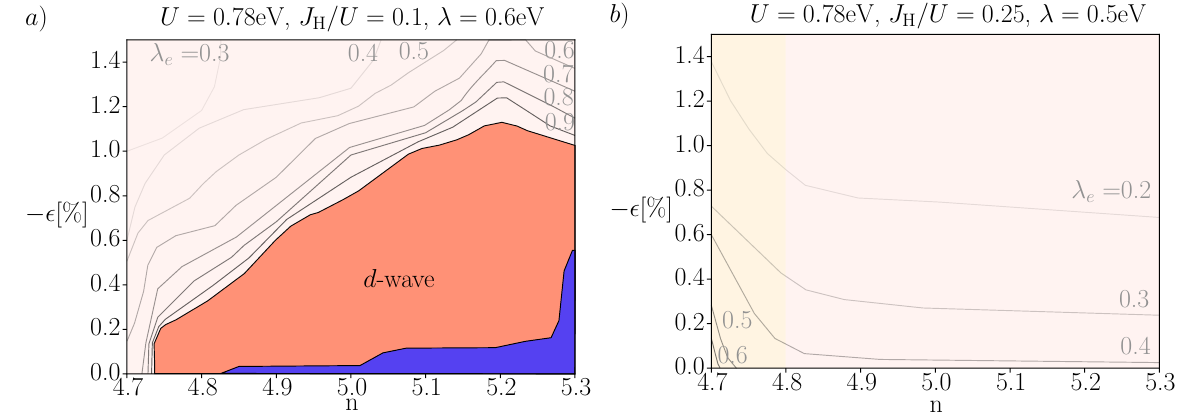} 
\caption{\label{fig:PDU078}For a choice of $U=0.78$ev$\approx 2.2 |t|$ the Stoner criterion is met only for some doping values, for the realistic interaction parameters in a). The superconducting region has a stronger doping-dependence than in Figs.~\ref{fig:PDr} \& \ref{fig:PDHund}. However, the strain values here are lower. In b), the largest eigenvalue is $\lambda_{e} <1$ for the higher Hund's coupling $J_{\textrm{H}}=0.25U$ and lower SOC $\lambda=0.5$eV. The pairing associated with the highest eigenvalue is an $s_{\pm}$-wave for $n < 4.8$ and a $d$-wave for all higher doping values. }
\end{figure*}

In the model used in this paper there is a trade-off between doping- and strain- dependent behavior for a given strength of the Hubbard interaction $U$, as motivated in section \ref{sec:PhaseDiagrams}.  The value for the calculated the phase diagram, $U \approx 3 |t|$, finds a magnetic phase transition for an expected range of strain values. In Fig.~\ref{fig:PDU078} phase diagrams are calculated at a lower $U=0.78$eV$\approx 2.2 |t|$. For the most realistic choice of Hund's ($J_{\textrm{H}}=0.1U$) and spin-orbit ($\lambda=0.6$eV), the magnetic region at $\epsilon=0$ only extends up to $n=4.85$ on the hole doped side. We note that for the chosen temperature and interaction parameters, a magnetic region extends beyond $n=5.3$. Mean field studies for the lattice with staggered rotations, such as in Appendix~\ref{sec:AppB}, reveal a possible canted ferromagnetic order for this doping. The high electron doping region is therefore likely to be a canted ferromagnet, as it has nesting vector $\boldsymbol{Q} = (\pi, \pi)$. The phase diagram has a clear doping-dependence of the superconducting region, with a dome centered at the electron doped regime. However, the considered strain values only extend up to $\epsilon=-1.5 \%$.

For the choice $J_{\textrm{H}}=0.25U$ and $\lambda=0.5$eV, neither a magnetic nor superconducting region is found for the temperature $T \approx 11$K. Similarly to the result at $U \approx 3 |t|$, the leading eigenvector corresponds to two different types of pairing, depending on the considered doping regime. 
\begin{figure*}
\centering   
\includegraphics[width=\textwidth]{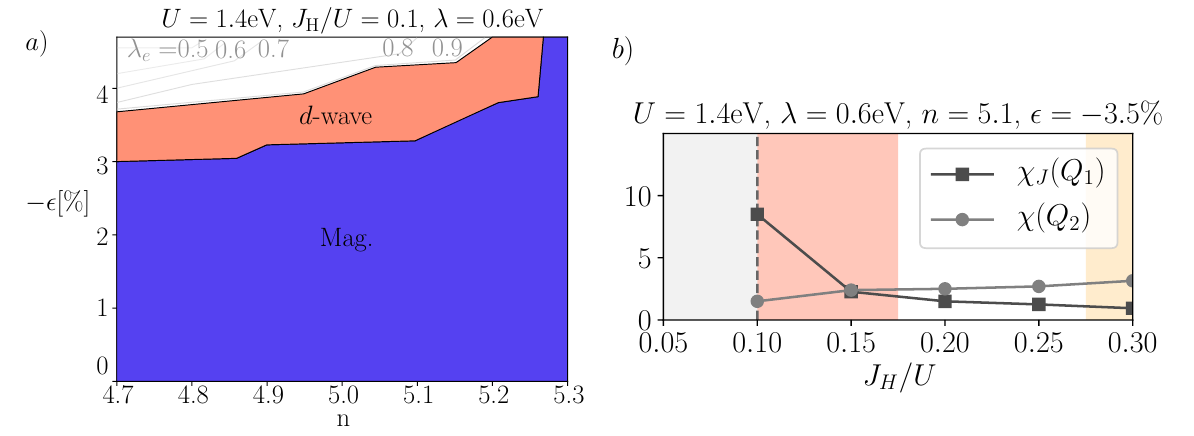} 
\caption{\label{fig:PDU14JH01}a) Phase diagram at $U=1.4$eV and additional parameters the same as in Fig.~\ref{fig:PDr}. The mainly $j=1/2$ $d$-wave order is present for a significant region in the higher strain regime. b) As in Fig.~\ref{fig:maxPeaks} the maximum values of pseudospin $\chi^{\uparrow \downarrow \uparrow \downarrow}_{J}(Q)$ and spin $\chi^{\uparrow \downarrow \uparrow \downarrow}(Q)$ susceptibility peaks are compared as the Hund's coupling is increased for one point in the phase diagram in a).} 
\end{figure*}
To explore the robustness of the strain-induces $d$-wave superconductivity, an additional phase diagram at $U \approx 4 |t|$ is shown in Fig.~\ref{fig:PDU14JH01}.
Even though the magnetic order persists up to higher values of the strain, the order is still a $d$-wave with mainly $j=1/2$ contributions. At higher strains, the pocket FS$_{2}$ is larger than in Fig.~\ref{fig:JHvarEigSymU11} and the ratio Max$\left[ \Delta_{\textrm{FS}_{2}}(\boldsymbol{k}) \right]/$Max$\left[ \Delta_{\textrm{FS}_{1}}(\boldsymbol{k}) \right] $ is increased.

\section{Mean field calculation} \label{sec:AppB}
The phase diagrams presented in this work are limited in the determination of competition between orders. Even though all superconducting orders are treated equally, they are calculated in the normal state. Therefore any influence of particle-hole self-energies are ignored. To compare phase boundaries for the regions of interest a self-consistent mean-field calculation was performed. We have chosen to only compare the RPA calculation for the most realistic phase diagram in Fig.~3. For the $j=1/2$ superconducting order we can make an approximation of the effectively attractive interaction between sites. However, the anisotropic $s$- and $d$-wave orders, as well as for the $p$-wave, require approximations of the effective attractive interactions between both nearest and next-nearest neighbor sites. As these multi-orbital pairing functions affect more than one $j$-state, one must also approximate the strength of the interaction in each of these channels. The mean field calculation was chosen to include all possible on-site magnetic order parameters, in a two-site unit cell, as well as for a $d$-wave superconducting order parameter in the $j=1/2$ state. The on-site order parameters, at sublattice $s=A,B$, are
\begin{equation}\label{eq:OPparam}
\begin{aligned}
\displaystyle & \langle c^{\dagger}_{s, a} c_{s, b} \rangle_{\textrm{MF},T} = \\ & \frac{1}{N_{k}^{2}} \sum_{n, \boldsymbol{k}} f(\xi_{\boldsymbol{k}, n}, T) \left[ U_{\textrm{MF}, \boldsymbol{k}}\right]_{(s,a) n} \left[ U^{\dagger}_{\textrm{MF}, \boldsymbol{k}}\right]_{n (s,b)} 
\end{aligned}
\end{equation}
with the orbit-spin label $a= (\alpha, \sigma)$, were calculated from a mean-field decoupling of the bare interactions in Eq.~\eqref{eq:Vbare}, as described in Ref.~\cite{Engstrom2021}.

The superconducting order parameter is introduced by mean field decoupling of the Bogoliubov-de Genne (BdG) Hamiltonian. $\Delta_{j=1/2}$ is set to be a $j=1/2$ $d$-wave singlet between the sublattices as
\begin{equation}
\begin{aligned}
   H_{\text{MF,SC}}  = \sum_{\boldsymbol{k}} & V(\epsilon)  \Delta_{j=1/2}  e^{i k_{x}} \left( \cos k_{x} - \cos k_{y}  \right) \\ & \times \left[  a_{\boldsymbol{k}, A, (\frac{1}{2}, + \frac{1}{2})} a_{-\boldsymbol{k}, B, (\frac{1}{2}, - \frac{1}{2})} \right. \\ & \left. -  a_{\boldsymbol{k}, A, (\frac{1}{2}, - \frac{1}{2})} a_{-\boldsymbol{k}, B, (\frac{1}{2}, + \frac{1}{2})} + \text{h.c.} \right]
  \end{aligned}
\end{equation}
where the operators $a$ are in the $j$-state basis, as in section~\ref{sec:ModelMethods}, and $V(\epsilon)=- \frac{3}{4} J_{\text{eff}}(\epsilon)$ is the effective interaction~\cite{Laughlin2014,Farrell2014}. Due to the structure of the hopping terms we approximate $J_{\text{eff}}(\epsilon)$ in the $j=1/2$ subspace as $J_{\text{eff}}(\epsilon)= \sqrt{J_{1}^{2} + D^2 }$~\cite{Seo2019}, where 
\begin{equation}
    J_{1} = \frac{4 \left(t_{\text{eff}}(\epsilon) - t_{\text{eff}, z}(\epsilon)\right)^2}{U_{\text{eff}}}, \qquad D = \frac{8 t_{\text{eff}}(\epsilon)  t_{\text{eff}, z}(\epsilon)}{U_{\text{eff}}}
\end{equation}
Here the effective hopping parameters are $t_{\text{eff}}(\epsilon) = \frac{1}{3} \left( t_{1} + t_{4} + t_{5} \right)$ and $t_{\text{eff}, z}(\epsilon) =  t_{rot} $. The order parameter is calculated via the self-consistency equation
\begin{equation}\label{eq:dOPparam}
\begin{aligned}
\displaystyle \Delta_{j=1/2} = \frac{1}{N_{k}^{2}} \sum_{n, \boldsymbol{k}} \sum_{\alpha, \beta} &e^{-i k_{x}} M_{\left(A, (\frac{1}{2}, + \frac{1}{2})\right), \alpha} M_{ \left(B, (\frac{1}{2}, - \frac{1}{2})\right), \beta} \\ & \times \left[ U_{\textrm{MF}, \boldsymbol{k}}\right]_{\alpha n} \left[ U_{\textrm{MF},\boldsymbol{k}}\right]_{\beta n}  f(\xi_{\boldsymbol{k}, n}, T)
\end{aligned}
\end{equation}
where the matrix $M$ is Eq.~\eqref{eq:Mmat}. This and Eq.~\eqref{eq:OPparam} are solved simultaneously via iterations as a set of coupled equations. Here we are treating only the interaction within the $j=1/2$ as if it was a one-band model, where the interaction projected onto that subspace is $H_{\text{eff}} = \left( U - \frac{4}{3} J_{\text{H}} \right) n_{i, (\frac{1}{2}, + \frac{1}{2})} n_{i, (\frac{1}{2}, - \frac{1}{2})} = U_{\text{eff}} n_{i, (\frac{1}{2}, + \frac{1}{2})} n_{i, (\frac{1}{2}, - \frac{1}{2})}$, so for the calculations $U_{\text{eff}}=\frac{13}{15} U$. An inclusion of $d$-wave pairing within the $(j,j_{z})=(\frac{3}{2}, \pm \frac{3}{2})$ sate or between the $j=3/2$ and $j=1/2$ does not extend the superconducting phase in the calculation. The interaction used here is $U \approx t_{\text{eff}}$, and not close to the strong coupling limit. However, to compare the competition between the $d$-wave and magnetic orders for for the RPA calculations results in an order of roughly equal size.
\begin{figure}
\centering   
\includegraphics[width=\linewidth]{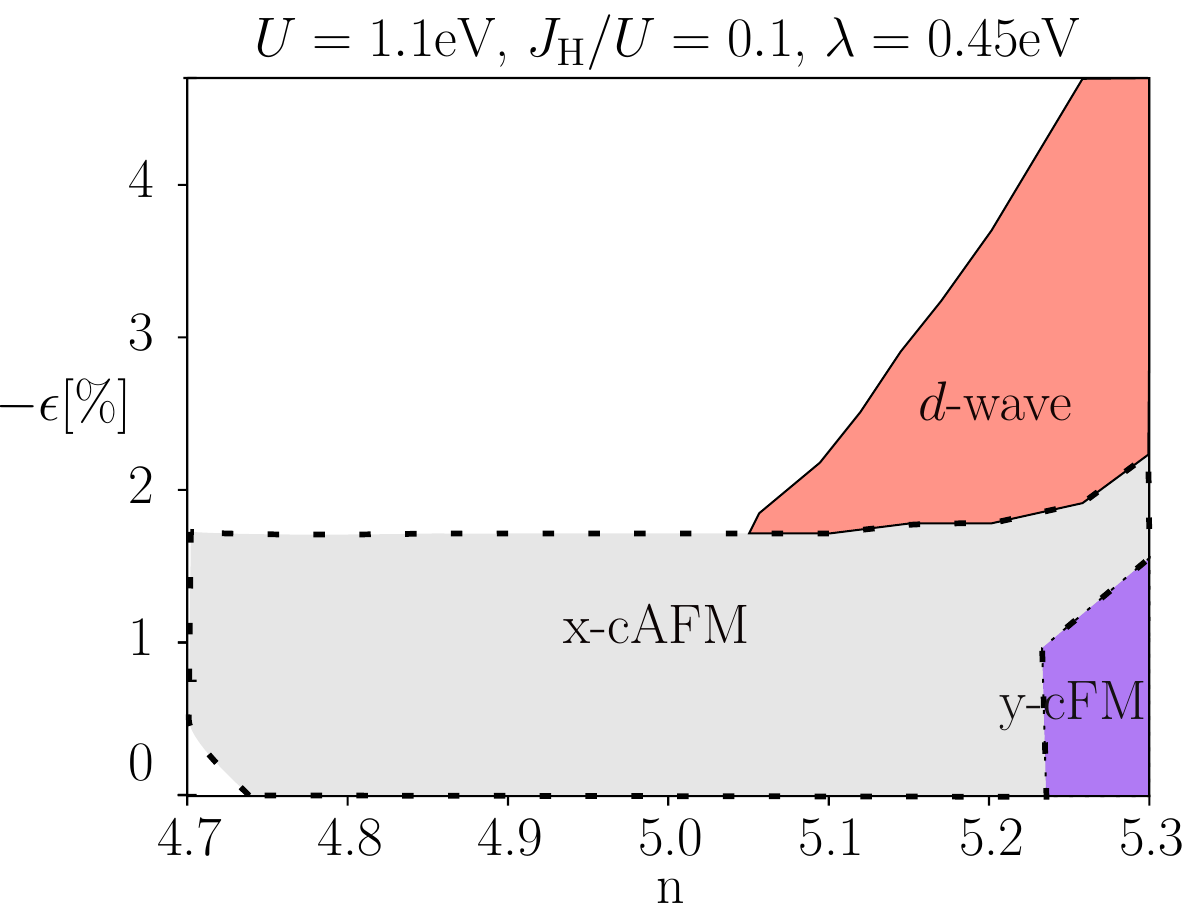} 
\caption{\label{fig:MFPD11JH01}The mean field calculation for magnetic and superconducting order parameters, for the same model parameters as in Fig.~\ref{fig:PDr} except $\lambda=0.45$eV. Here $V=-0.25U \approx - \frac{3}{4} J_{\text{eff}}(- 2.5 \%)$. As the SOC is renormalized the resulting bandstructure is the same as for the phase diagram in the main text.}
\end{figure}

Calculations were performed on a $N_{k} \times N_{k}$ lattice in momentum space, with $N_{k}=100$, and self-consistent solutions were found iteratively with a convergence criterion of $10^{-7}$. Due to the full set of order parameters containing terms that renormalize the spin-orbit coupling, the self-consistent mean field calculation predicts effects not included in the RPA calculation. However, the included mean field orders can only have nesting vectors $\boldsymbol{Q}=(0,0)$ or $\boldsymbol{Q}=(\pi,\pi)$ as the two sites allow us to study either net $(\langle c^{\dagger}_{A, a} c_{A, b} \rangle_{\textrm{MF},T} + \langle c^{\dagger}_{B, a} c_{B, b} \rangle_{\textrm{MF},T} )/2$ or staggered $(\langle c^{\dagger}_{A, a} c_{A, b} \rangle_{\textrm{MF},T} - \langle c^{\dagger}_{B, a} c_{B, b} \rangle_{\textrm{MF},T} )/2$ values of the order parameters.

In Fig.~\ref{fig:MFPD11JH01} there are two magnetic mean field orders. The most common order is the canted in-plane antiferromagnet (x-cAFM), with a staggered magnetic moment along the $x$-direction and a net magnetic moment along the $y$-direction. Similarly to previous works, the canting angle follows the rotations of the octahedra in the lattice. When a compressive strain increases the rotations, the magnetic canting angle thus increases as well. For a high enough $U$ the magnetic order remains up to high electron number ($n>5.2$) the in-plane ferromagnetic component becomes dominant. A small AFM component remains, identifying this order as y-cFM. The canted orders consist mainly of $j=1/2$ pseudospins. However, as the SOC is lowered the AFM order in the hole doped region has small $(j,j_{z})=(\frac{3}{2}, \pm \frac{3}{2})$ contributions. The order parameters which renormalize the SOC depend on strain and doping. However, if $\lambda=0.45$eV is chosen the effective spin orbit coupling $\lambda_{\text{eff}} \approx 0.6$eV.

A superconducting region is present in Fig.~\ref{fig:MFPD11JH01} and the competition between superconductivity and magnetism therefore does not affect its existence. However, in contrast to Fig.~\ref{fig:PDr} superconductivity is only present for electron doping $n>5.05$. For the non-interacting band structure, used for the RPA calculations in this paper, the $j=3/2$ pocket is present in the full charge doping region around $\epsilon=-2.2\%$. As the $j=3/2$ hole pocket increases in size for a given charge doping, the $j=1/2$ electron pocket grows as well. The superconductivity in Fig.~\ref{fig:PDr} is therefore present for a region where the $j=1/2$ electron pocket is significantly larger than for $\epsilon=0$. There the $d$-wave is present for all considered $n$. 

The discrepancy between the RPA and the mean field result is due to several factors. Since these two approaches make different approximations it is not possible to determine which phase diagram is more realistic. Instead, we can gain confidence in our results in parts of the phase diagram where the two approaches agree. Each approach has its strength and weaknesses. In mean field we must pre-determine the possible channels of superconductivity and the effective attractive interaction which does not change with strain and doping. On the other hand, the self-consistency equation Eq.~\eqref{eq:dOPparam} is not linearized like the RPA gap equation in Eq.~\eqref{eq:linGap} and therefore the mean field is better suited for determining the relative strength of the order parameters considered. The comparison between RPA and mean field theory therefore suggest that the electron doped region is more likely to host a d-wave superconducting order.
\section{Tetragonal splitting} \label{sec:AppC}
\begin{figure}
\centering    
\includegraphics[width=0.9\linewidth]{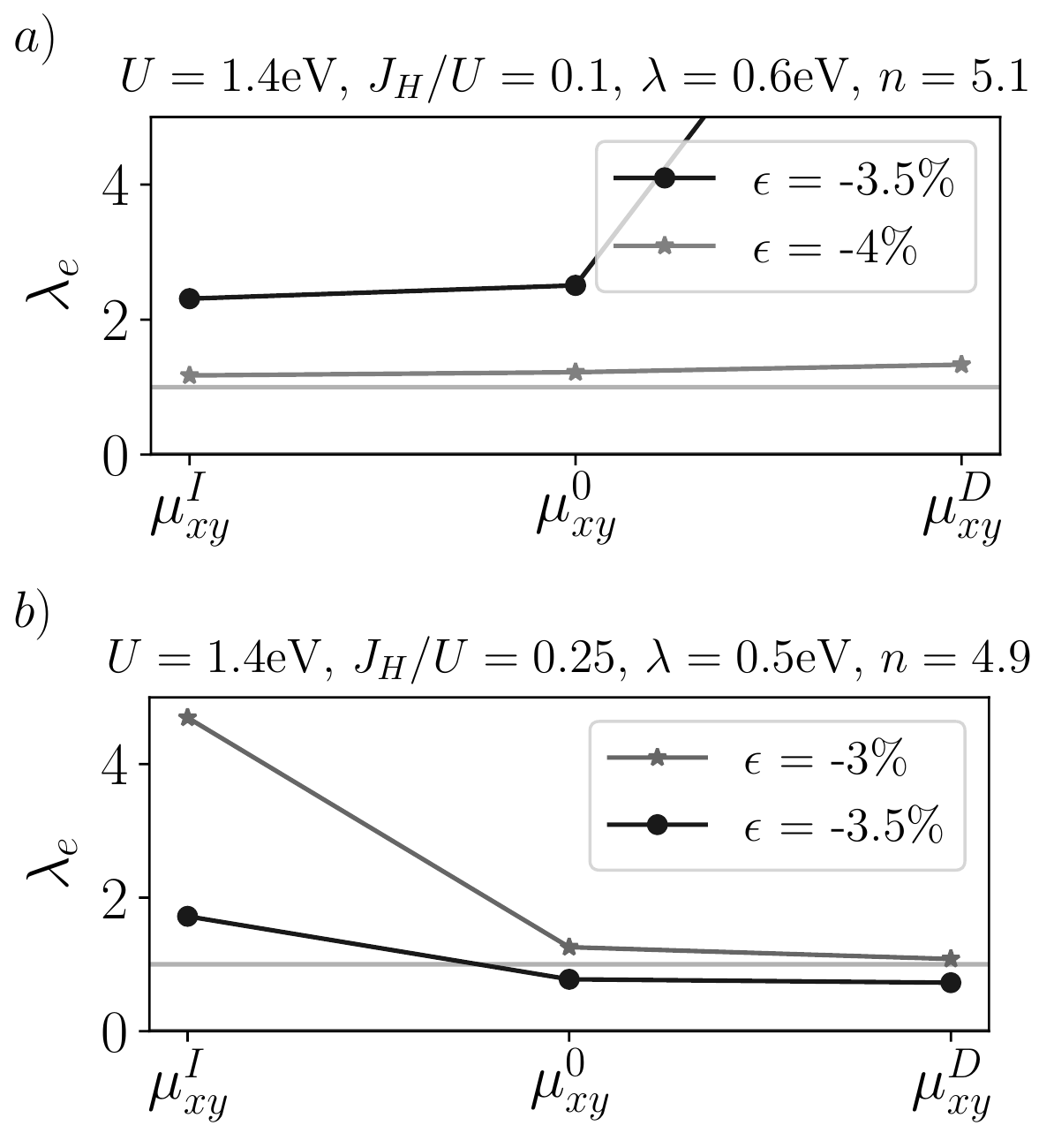} 
\caption{\label{fig:eigmuxy}The largest eigenvalue $\lambda_{e}$ is compared for the three options for the strain-dependence of $\mu_{xy}(\epsilon)$, as given by Eqs.~\eqref{eq:muI} \& \eqref{eq:muD}. For $J_{\textrm{H}}=0.1U$ and SOC $\lambda=0.6$eV we note the increasing eigenvalue, for a $d$-wave order, as the absolute value of $\mu_{xy}$ decreases. For a lower $|\mu_{xy}|$, the bands of $(3/2, \pm 3/2)$-character are pushed further down and a larger fraction of bands at the Fermi surface has $j=1/2$ character. For $J_{\textrm{H}}=0.25U$ and lower SOC, the opposite trend is observed and the order is the $s_{\pm}$-wave.}
\end{figure}
\begin{figure*}
\centering   
\includegraphics[width=\textwidth]{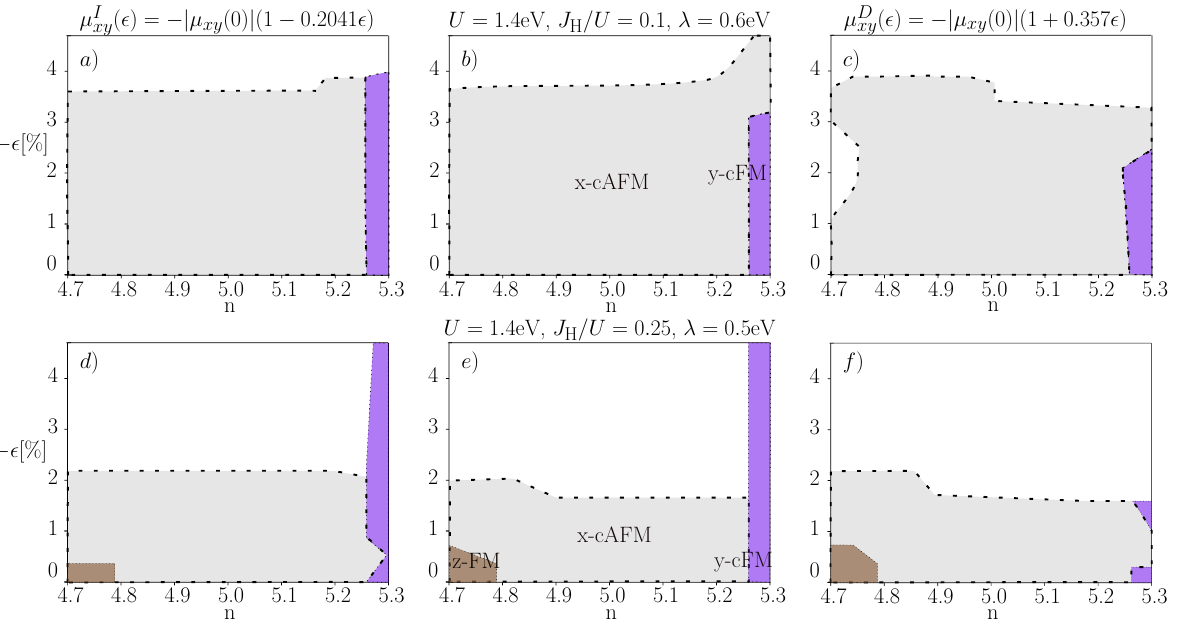} 
\caption{\label{fig:muxyPDJH01}The magnetic mean field phase diagrams, Eq.~\eqref{eq:OPparam}, for three different options for $\mu_{xy}(\epsilon)$: $\mu_{xy}^{I}$ (absolute value increasing, Eq.~\eqref{eq:muI}), $\mu_{xy}^{0}$ (constant), $\mu_{xy}^{D}$ (absolute value decreasing, Eq.~\eqref{eq:muD}). No significant shift of the amount of strain required for a phase transition is observed. A larger absolute value of the splitting favors a magnetic order for electron doping. At the strain values required for the transition, $\mu_{xy}^{D}$ has changed sign, $\mu_{xy}^{D}>0$, and the order instead favors hole doping. For $J_{\textrm{H}}=0.25U$ a lower splitting favors the z-FM order while it suppresses the y-cFM order.}
\end{figure*}
\begin{figure*}
\centering   
\includegraphics[width=\textwidth]{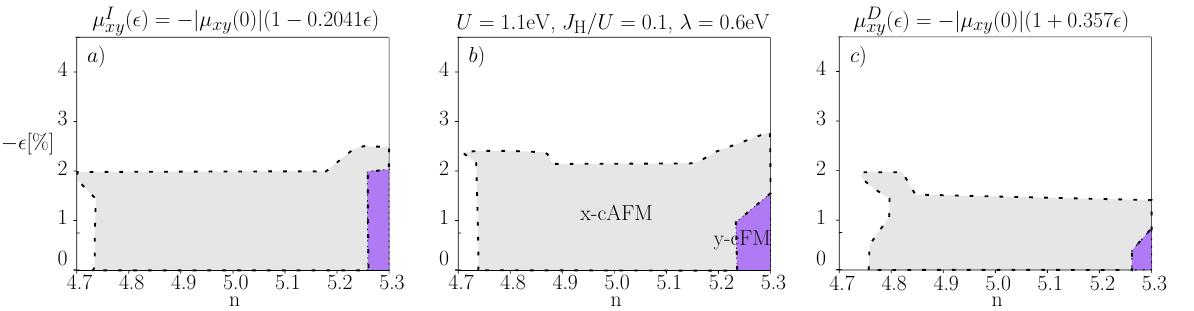} 
\caption{\label{fig:muxyPDJH01U11}The magnetic mean field phase diagrams at a lower $U=1.1$eV. Since the magnetic order does not remain up to as high strains, the tetragonal splitting effects are less prominent. Only $J_{\textrm{H}}=0.1U$ is shown here as the magnetic regions for $J_{\textrm{H}}=0.25U$ are too small to see any difference between the three models.}
\end{figure*}
Compression has an additional effect relevant to the iridates: an increased tetragonal distortion. The tetragonal distortion of the oxygen octahedra encompassing the iridium atoms has been measured to increase with compression - becoming more elongated when the in-plane compression increases. For unstrained Sr$_{2}$IrO$_{4}$ a small elongation is observed, which theoretically should result in a tetragonal splitting $\mu_{xy}>0$. However, early \textit{ab initio} calculations found that the band structure is best described by a shift $\mu_{xy}<0$~\cite{Kim2008,Zhang2013}. Later works have proposed that the sign could arise from hybridization with ligand oxygen orbitals~\cite{Agrestini2017,Lane2020}. Due to this sign difference, previous works modeling the tetragonal splitting's dependence on strain come to contradictory results, where $|\mu_{xy}|$ either increases or decreases~\cite{Jackeli2009,Zhang2013, Bhandari2019}. A fitting of the change in energy splitting to RIXS measurements, in Ref.~\cite{Paris2020}, found an increasing $|\mu_{xy}|$ for low compressive strain. A linearization of these results gives:
\begin{equation}\label{eq:muI}
\mu_{xy}^{I}(\epsilon) = -|\mu_{xy}(0)| (1 -0.2041\epsilon). 
\end{equation}
where $\epsilon$ is given in units of \%. In Fig.~\ref{fig:eigmuxy} the largest eigenvalue for the linearized gap equation is compared for an approximation of $\mu_{xy}(\epsilon)$ as given by either Ref.~\cite{Paris2020} or Ref.~\cite{Bhandari2019}. The second approximation is based on theoretical calculations and the linearization is instead
\begin{equation}\label{eq:muD}
\mu_{xy}^{D}(\epsilon)= -|\mu_{xy}(0)| (1 + 0.357 \epsilon) 
\end{equation}
which results in a decreasing $|\mu_{xy}|$ under strain. We can note that even though the experimentally approximated values are only based on data points up to $\epsilon = -0.7$\%, the theoretical approximation $\mu_{xy}^{D}(\epsilon)$ is not compatible with the found trend. The experimentally motivated $\mu_{xy}^{I}(\epsilon)$ results in a slightly lower eigenvalue than the constant $\mu_{xy}$ for the $d$-wave order, at $\lambda=0.6$eV and $J_{\textrm{H}}=0.1U$. Any change to the tetragonal splitting is thus expected to have a small impact on the strain-induced superconducting regions with $d$-wave symmetry.

As a check of the magnetic phase boundaries, for different values of the tetragonal splitting, the mean field calculation in Appendix~\ref{sec:AppA} was performed, for only the magnetic order parameters, at $N_{k}=200$. In Fig.~\ref{fig:muxyPDJH01}, an additional out-of-plane ferromagnetic order (z-FM) appears at the lower SOC and higher Hund's coupling. This order is only favored at high enough $U$ and has a clear spin character, $\langle L_{z} \rangle/ \langle S_{z} \rangle \approx 0.16$. Contributions from each orbital are of approximately equal strength. In terms of $j$-states the contributions are thus mainly from the $(j,j_{z})=(\frac{1}{2}, \pm \frac{1}{2})$ and $(j,j_{z})=(\frac{3}{2}, \pm \frac{3}{2})$ states.

As seen in Figs.~\ref{fig:muxyPDJH01} \& \ref{fig:muxyPDJH01U11}, a larger absolute value of the splitting, $\mu_{xy}^{I}(\epsilon)$, does not result in major changes to the magnetic phase boundaries. For the other option, $\mu_{xy}^{D}(\epsilon)$, the behavior with doping changes as $\mu_{xy}^{D}(\epsilon \approx -2.8 \%)=0$. The x-cAFM order remains up to higher strains for hole instead of electron doping once the sign of $\mu_{xy}$ changes.

\section{Symmetry classification}\label{sec:AppD}
\begin{figure*}
\centering    
\includegraphics[width=\textwidth]{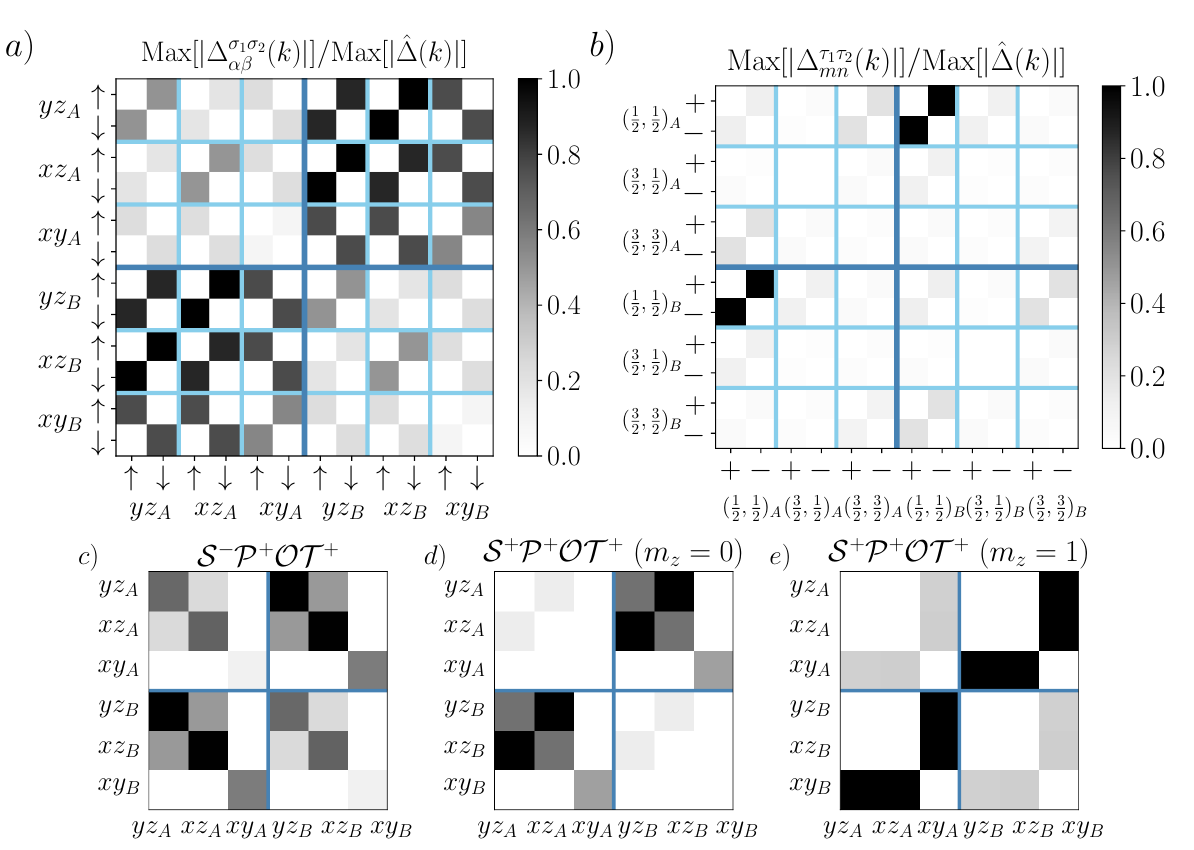} 
\caption{\label{fig:dSym}The calculated even parity $d$-wave pairing (found eigenstate to the largest eigenvalue at the point $U=1.1$eV, $J_{\textrm{H}}=0.1U$, $\lambda=0.6$eV, $n=5.1$, $\epsilon=-2\%$), where the maximum value of each component is shown both in a) the orbital $\Delta_{\alpha \beta}^{\sigma_{1} \sigma_{2}}$ and in the b) $j$-state bases $\Delta_{m n}^{\tau_{1} \tau_{2}}$. The largest value of the order parameter is normalized to Max$|\hat{\Delta}(k)|=1$. In the orbital basis, the pairing has many inter-sublattice components of equal size. As shown in c),d),e), considering the pairing with even or odd spin symmetries results in both large spin-singlet and spin-triplet components. In $j$-state components the pairing is clearly dominated by a $j=1/2$ pseudospin singlet.}
\end{figure*}
\begin{figure*}
\centering    
\includegraphics[width=\textwidth]{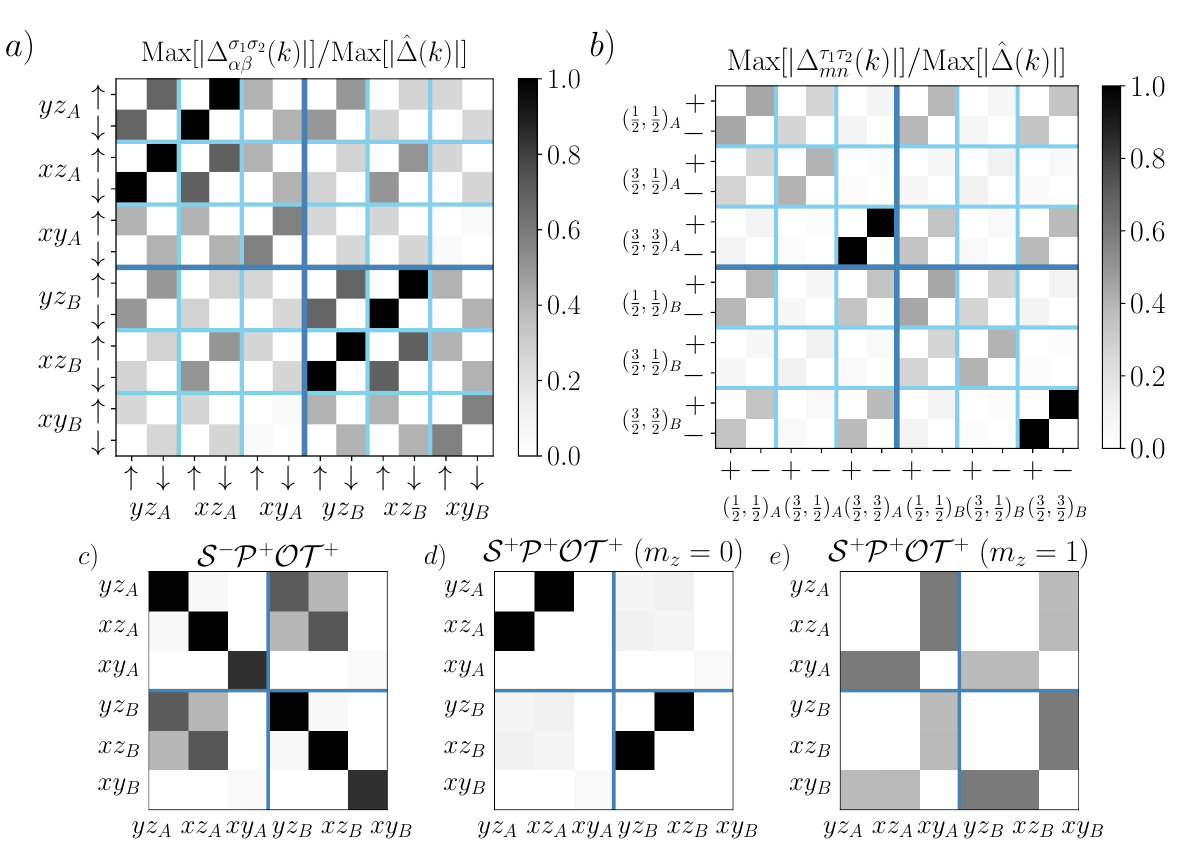} 
\caption{\label{fig:spmSym}The calculated even parity $s_{\pm}$-wave pairing (at $U=1.1$eV, $J_{\textrm{H}}=0.3U$, $\lambda=0.5$eV, $n=4.8$, $\epsilon=-1\%$). This order is a combination of several components both in a) the orbital and b) $j$-state bases, which are normalized by the largest value of the order parameter Max$|\hat{\Delta}(k)|$. The pairing consists mainly of intra-sublattice pseudospin-singlets. The strongest components at this point in the phase diagram in the $(j,j_{z})=(\frac{3}{2}, \pm \frac{3}{2})$-state, with a $(j,j_{z})=(\frac{1}{2}, \pm \frac{1}{2})$ pseudospin singlet following in size.}
\end{figure*}
In section \ref{sec:Symmetries} the $\mathcal{S}\mathcal{P}\mathcal{O}\mathcal{T}$-formalism~\cite{Sigrist1991,Linder2019} of classifying the symmetry of the superconducting pairing is introduced. The pairing must be antisymmetric under the product of operators
\begin{equation}
    \mathcal{S}\mathcal{P}\mathcal{O}\mathcal{T} \Delta_{ab}(k) = \Delta_{ba}(-k) 
\end{equation}
where each exchange operator is defined as
\begin{equation}
\begin{aligned}
    \mathcal{S} \Delta_{(s, \alpha, \sigma) (s', \alpha', \sigma')}(\boldsymbol{k}, \omega) =& \Delta_{(s, \alpha, \sigma') (s', \alpha', \sigma)}(\boldsymbol{k}, \omega) \\ \mathcal{P} \Delta_{(s, \alpha, \sigma) (s', \alpha', \sigma')}(\boldsymbol{k}, \omega) =& \Delta_{(s, \alpha, \sigma) (s', \alpha', \sigma')}(-\boldsymbol{k}, \omega) \\
    \mathcal{O} \Delta_{(s, \alpha, \sigma) (s', \alpha', \sigma')}(\boldsymbol{k}, \omega) =& \Delta_{(s, \alpha', \sigma) (s', \alpha, \sigma')}(\boldsymbol{k}, \omega) \\ \mathcal{T} \Delta_{(s, \alpha, \sigma) (s', \alpha', \sigma')}(\boldsymbol{k}, \omega) =& \Delta_{(s, \alpha, \sigma) (s', \alpha', \sigma')}(\boldsymbol{k}, -\omega) 
    \end{aligned}
\end{equation}
In Fig.~\ref{fig:dSym} the symmetries of a found $j=1/2$ $d$-wave pairing is shown. The maxima of each component of $\Delta$ are separated under the present spin-singlet and spin-triplet operations. An ideal pseudospin singlet will have both spin-singlet and spin-triplet components of comparable size, in the orbital-spin basis. As the transformation between bases is independent of momentum, the spatial parity of the state is unchanged. The spin-triplet components are therefore orbital-singlets $\mathcal{S}^{+} \mathcal{P}^{+} \mathcal{O}^{-}\mathcal{T}^{+}$. The found $d$-wave in Fig.~\ref{fig:dSym} has additional small components from other $j$-states. The compressive strain decreases the contributions from the $xy$-orbital at the Fermi surface as well as increases the interorbital $xz$-$yz$ hopping. There are therefore more contributions from spin-singlet and $m_{z}=0$ spin-triplet components, than from $m_{z}=1$ triplets. In Fig.~\ref{fig:spmSym} the symmetries of an $s_{\pm}$-wave is shown. The $s_{\pm}$-wave has strong components from several $j$-states. As this order is found to be mediated by spin-like fluctuations mainly in the $yz$- and $xz$-orbitals, between bands of $j=1/2$ and $j=3/2$ character, we can consider the main components of the pairing to come from pairing within and between those orbitals. 

The superconducting order can be expressed exactly via its full symmetry representation. We consider the pairing for a spin and orbital combination:
\begin{equation}
    \Delta_{\sigma \sigma'}^{\alpha \beta} (\boldsymbol{k}) = \frac{1}{2} \sum_{s,s'} e^{i k_{x} (\Theta_{s} - \Theta_{s'} )}  \Delta_{\sigma \sigma'}^{(\alpha,s) (\beta,s')} (\boldsymbol{k}) 
\end{equation}
where $\Theta_{s}$ is the same function as in Eq.~\eqref{eq:ChiJtot} and gives us the combined contribution from both sublattice sites. The symmetry representation for the spatial symmetry $\eta^{\mu} (\boldsymbol{k})$ is specified in Table~\ref{tab:SymIrr}. The pairing can be decomposed into symmetry representations for the spin and orbital structure. If $\mathcal{C}_{\mu \nu \rho}$ is the projection constant for a chosen representation, then any superconducting order can be written as:
\begin{equation}
    \Delta_{\sigma \sigma'}^{\alpha \beta} (\boldsymbol{k}) = \sum_{\mu,\nu, \rho} \mathcal{C}_{\mu \nu \rho} \eta^{\mu} (\boldsymbol{k}) S^{\nu}_{\sigma \sigma'} O^{\rho}_{\alpha \beta}
\end{equation}
For the spin degree of freedom, $S^{\nu}_{\sigma \sigma'}$ are the generators for the SU(2) algebra, the Pauli matrices $\sigma^{i}$ with $i=0, x,y,z$. For the orbital degree of freedom, $O^{\rho}_{\alpha \beta}$ are the generators for the SU(3) algebra, the Gell-Mann matrices~\cite{GellMann1961} $\lambda_{i}$ with $i=0, 1,\dots,8$. The Pauli matrices act in spin $(\uparrow, \downarrow)$ space and can form spin-triplets $(\sigma^{0}, \sigma^{x}, \sigma^{z})$ and spin-singlets $(\sigma^{y})$. The Gell-Mann matrices act in orbital $(d_{yz}, d_{xz}, d_{xy})$ space and can form intraorbital pairings $(\lambda_{0},\lambda_{3},\lambda_{8})$, and interorbital pairings that can be either even $(\lambda_{1},\lambda_{4},\lambda_{6})$ or odd $(\lambda_{2},\lambda_{5},\lambda_{7})$ under orbital exchange.

The found $d$-wave is a pseudospin singlet, which within the $j=1/2$ subspace is
\begin{equation}\label{eq:dSym}
    \Delta_{d}(\boldsymbol{k}) \approx \eta^{B_{1g}}_{R=1} (\boldsymbol{k})\otimes  (i \tilde{\sigma}^{y} )
\end{equation}
where $\tilde{\sigma}^{y}$ acts on the pseudospins $\tau=+,-$. The $s_{\pm}$-wave can be expressed approximately as intraorbital spin-singlets and interorbital spin-triplets in the $yz$- and $xz$-orbitals:
\begin{equation}
\begin{aligned}\label{eq:spmSym}
    \Delta_{s_{\pm}}(\boldsymbol{k}) \approx & \left( \mathcal{C}^{A_{1g}}_{R=0} \eta^{A_{1g}}_{R=0} (\boldsymbol{k}) + \eta^{A_{1g}}_{R=2} (\boldsymbol{k})  \right) \otimes \\  &  \left[ (i \sigma^{y}) \otimes \left(  \lambda_{11} + \lambda_{22}  \right) -i  \sigma^{x} \otimes \left( i \lambda_{2} \right) \right]
    \end{aligned}
\end{equation}
with $\lambda_{11}= \left( \frac{1}{3} \lambda_{0} + \frac{1}{2} \lambda_{3}  + \frac{1}{2 \sqrt{3}} \lambda_{8} \right) $ and $\lambda_{22}= \left( \frac{1}{3} \lambda_{0} - \frac{1}{2} \lambda_{3}  + \frac{1}{2 \sqrt{3}} \lambda_{8} \right) $ representing intraorbital pairing. The A$_{1g}$ spatial symmetry has two contributions with a relative weight specified by $\mathcal{C}^{A_{1g}}_{R=0} \approx 0.7$. This pairing has additional smaller contributions involving the $xy$-orbital: $\propto  (i \sigma^{y}) \otimes \lambda_{33} = (i \sigma^{y}) \otimes \left( \lambda_{0} -  \sqrt{3} \lambda_{8} \right)$, $\propto (-i) \sigma^{0} \otimes \left( i \lambda_{5} \right)$, and $\propto (-i) \sigma^{z} \otimes \left( i \lambda_{7} \right)$.

\section{Anisotropic pairing}\label{sec:AppE}
\begin{figure*}
\centering   
\includegraphics[width=\textwidth]{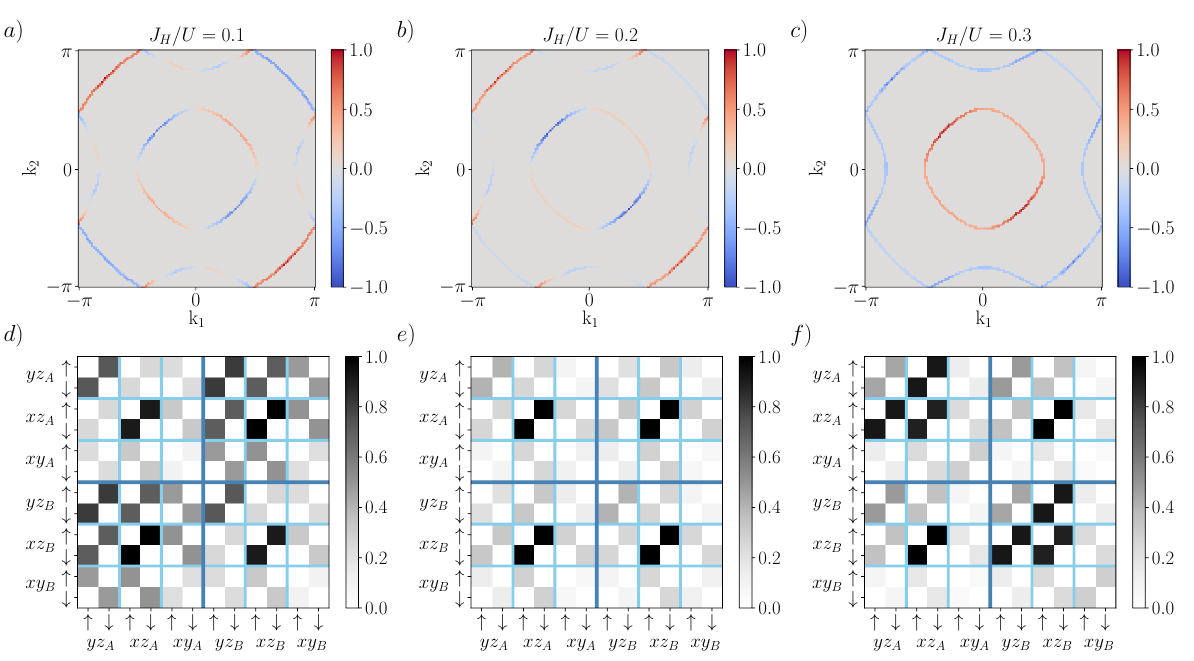} 
\caption{\label{fig:FSprojxz}a),b),c): The intraband pairing for the anisotropic superconducting orders is shown on the FS as the Hund's coupling $J_{\textrm{H}}$ is increased, for the same values as in Fig.~\ref{fig:JHvarEigSymla05}. All orders are a mix of $s$- and $d$-wave symmetries, with only the one at $J_{\textrm{H}}=0.3U$ being node-less. d), e),f): The maximum of pairing components in the orbital basis for the same Hund's coupling as the plot above, shown as $\textrm{Max}\left[|\Delta_{\alpha \beta}^{\sigma \sigma'} (\boldsymbol{k})| \right] / \textrm{Max}\left[|\hat{\Delta} (\boldsymbol{k})| \right]$. For all values the order exists mainly in the $yz$- and $xz$-orbitals, with barely any contributions from $xy$. However, the order is stronger in $xz$. At $J_{\textrm{H}}=0.2U$ the pairing originates almost entirely from the $xz$-orbital.}
\end{figure*}
The anisotropic orders found at high compressive strain, in section \ref{sec:Hunds}, appear when fluctuations with $\boldsymbol{Q}$-vectors which connect states of the same orbitals become large enough. In Fig.~\ref{fig:FSprojxz} the found anisotropic orders, for three values of the Hund's coupling, are characterized by projecting the pairing on the Fermi surface as well as the maximal value of each orbital-spin component of the pairing. At the Fermi surface the weight is stronger along the $x$-direction, on all pockets. The $xz$ orbital has the largest contribution. The other notable difference, from the $d$- and $s_{\pm}$-wave orders, is that the anisotropic orders have equal size intra- and inter-sublattice components of the $xz$-orbital. 

The anisotropic orders and $s_{\pm}$-order are mediated mainly by spin fluctuations. However, the pockets at the Fermi surface have a strong character of the $(j,j_{z})=(\frac{1}{2}, \pm \frac{1}{2})$ or $(j,j_{z})=(\frac{3}{2}, \pm \frac{3}{2})$ state. The orders originate mostly from the $yz$- and $xz$-orbitals, and the pairing can be transformed into the $j$-states, via Eq.~\eqref{eq:Mmat}, as
\begin{equation}
\begin{aligned}
    \Delta^{\tau \tau'}_{mn}(k) = \sum_{\alpha \beta \sigma \sigma'} c^{(\tau \tau')(\sigma \sigma')}_{(mn)(\alpha \beta)} \Delta^{\sigma \sigma'}_{\alpha \beta}(k) \\ =  \sum_{\alpha \beta \sigma \sigma'} M_{(m,\tau), (\alpha, \sigma)} M_{(n,\tau'), (\beta, \sigma')} \Delta^{\sigma \sigma'}_{\alpha \beta} (k)
\end{aligned} 
\end{equation}
where $m=\left( \frac{1}{2}, \pm \frac{1}{2} \right), \left( \frac{3}{2}, \pm \frac{1}{2} \right), \left( \frac{3}{2}, \pm \frac{3}{2} \right)$, $\tau=+,-$, $\alpha=yz,xz,xy$, and $\sigma=\uparrow, \downarrow$. We can identify $c^{(+-)(\uparrow \downarrow)}_{(\frac{3}{2}, \pm \frac{3}{2}) (yz,yz)} =  c^{(+-)(\uparrow \downarrow)}_{(\frac{3}{2}, \pm \frac{3}{2}) (xz,xz)} = - \frac{1}{2}$ and $c^{(+-)(\uparrow \downarrow)}_{(\frac{1}{2}, \pm \frac{1}{2}) (yz,yz)} =  c^{(+-)(\uparrow \downarrow)}_{(\frac{1}{2}, \pm \frac{1}{2}) (xz,xz)} = \frac{1}{3}$. As can be seen in Fig.~\ref{fig:Q1Q2Peaks}, sections of the hole doped Fermi surface are mainly of either $yz$- or $xz$-character. This is a result of the quasi-1d dispersion in each of these orbitals in Eq.~\eqref{eq:kinTerms}. A simplified model for the two Fermi surfaces is to introduce orbitals with a spatial dependence along the parameter $\theta$ around a circular FS:
\begin{equation}
\begin{aligned}
| yz \rangle_{\theta} =& |\cos \theta| | yz \rangle \\ | xz \rangle_{\theta} =& |\sin \theta|  | xz \rangle \\ | xy \rangle_{\theta} =& \left(|\cos \theta| +  |\sin \theta| 
 \right) |xy \rangle
\end{aligned}
\end{equation}
Here only one site and the full Brillouin zone (BZ) are considered for simplicity. So in a BZ where FS$_{1}$ is purely $(\frac{1}{2}, \pm \frac{1}{2})$ and FS$_{2}$ is $(\frac{3}{2}, \pm \frac{3}{2})$ (with an energy shift $\xi$ of the $xy$-orbital):
\begin{equation}
\begin{aligned}
|\textrm{FS}_{1}, \uparrow \rangle_{\theta} = |\frac{1}{2}, + \frac{1}{2} \rangle_{\theta} - \xi | xy, \uparrow \rangle_{\theta}  \\
 =\frac{1}{\mathcal{N}_{1}}\left( | yz, \downarrow \rangle_{\theta} -i | xz, \downarrow \rangle_{\theta}  + (1- \sqrt{3} \xi) | xy, \uparrow \rangle_{\theta} \right)
 \end{aligned}
\end{equation}
and
\begin{equation}
|\textrm{FS}_{2}, \uparrow \rangle_{\theta} = |\frac{3}{2}, + \frac{3}{2} \rangle_{\theta} =\frac{1}{\mathcal{N}_{2}}\left( -| yz, \downarrow \rangle_{\theta} -i | xz, \downarrow \rangle_{\theta} \right) 
\end{equation}
where $\mathcal{N}_{1},\mathcal{N}_{2}$ are normalization factors. For a simplified constant order only within each orbital, $\Delta^{\uparrow \downarrow}_{yz, yz}(k)=\Delta$ the pairing on each FS becomes 
\begin{equation}
\Delta^{\downarrow \uparrow}_{\textrm{FS}_{1}} = \frac{\cos^{2} \theta}{\mathcal{N}_{1}^{2}}  \Delta, \qquad \Delta^{\downarrow \uparrow}_{\textrm{FS}_{2}}= - \frac{\cos^{2} \theta}{\mathcal{N}_{2}^{2}} \Delta
\end{equation}
If instead $\Delta^{\uparrow \downarrow}_{xz,xz}=\Delta$
\begin{equation}
\Delta^{\downarrow \uparrow}_{\textrm{FS}_{1}} = \frac{\sin^{2} \theta}{\mathcal{N}_{1}^{2}}  \Delta, \qquad \Delta^{\downarrow \uparrow}_{\textrm{FS}_{2}}= - \frac{\sin^{2} \theta}{\mathcal{N}_{2}^{2}} \Delta
\end{equation}
Each orbital thus results in a quasi-1d pairing on both Fermi surfaces. If one of the orbitals dominate $\Delta^{\uparrow \downarrow}_{xz,xz} > \Delta^{\uparrow \downarrow}_{yz,yz}$ the order is an anisotropic $s$-wave, with some $d$-wave components.

To study the full $s_{\pm}$-wave, it can be modeled as equal parts from both orbitals, $\Delta^{\uparrow \downarrow}_{yz,yz}=\Delta^{\uparrow \downarrow}_{xz,xz}=\Delta$:
\begin{equation}
\begin{aligned}
\Delta^{\downarrow \uparrow}_{\textrm{FS}_{1}} =& \frac{\cos^{2} \theta + \sin^{2} \theta}{\mathcal{N}_{1}^{2}}  \Delta &= \frac{1}{\mathcal{N}_{1}^{2}}  \Delta \\\Delta^{\downarrow \uparrow}_{\textrm{FS}_{2}}=& - \frac{\cos^{2} \theta + \sin^{2} \theta}{\mathcal{N}_{2}^{2}} \Delta &=  - \frac{1}{\mathcal{N}_{2}^{2}} \Delta
\end{aligned}
\end{equation}
where in the ideal $j$-state case $\mathcal{N}_{1}^{2}=3$ and $\mathcal{N}_{2}^{2}=2$. This is one of the primary reasons for the $s$-wave symmetry resulting in an $s_{\pm}$-wave with a larger weight on FS$_{2}$. However, the found $s_{\pm}$-wave does not have purely $\mathcal{S}^{-}\mathcal{P}^{+}\mathcal{O}^{+}\mathcal{T}^{+}$ intraorbital components but also interorbital $\mathcal{S}^{+}\mathcal{P}^{+}\mathcal{O}^{-}\mathcal{T}^{+}$-terms, see Fig.~\ref{fig:spmSym}. As these terms all have the same magnitude $\Delta^{\uparrow \downarrow}_{yz,xz}= - \Delta^{\uparrow \downarrow}_{xz,yz}= i\Delta$. Projected onto the Fermi surface
\begin{equation}
\Delta^{\downarrow \uparrow}_{\textrm{FS}_{1}} = \frac{2 |\cos \theta | | \sin \theta |}{\mathcal{N}_{1}^{2}}  \Delta, \qquad \Delta^{\downarrow \uparrow}_{\textrm{FS}_{2}}= \frac{2 |\cos \theta | | \sin \theta |}{\mathcal{N}_{2}^{2}} \Delta.
\end{equation}
They contribute to both bands with the same sign and to the same sections. The pairing used for these examples so far has been a uniform $s$-wave, $\Delta(\boldsymbol{k}) = \Delta$. In Eq.~\eqref{eq:spmSym} we can note that the found $s_{\pm}$-wave has a dependence on momentum, with a large contribution from the $\eta^{A_{1g}}_{R=2} (\boldsymbol{k})= 2 \cos k_{x} \cos k_{y}$ symmetry. The placement of FS$_{1}$ and FS$_{2}$ in the BZ therefore affects the sign of the gaps. As a result, both intra- and inter-orbital terms play a role in the origin of the $s_{\pm}$-wave pairing.
\begin{figure}
\centering    
\includegraphics[width=0.83\linewidth]{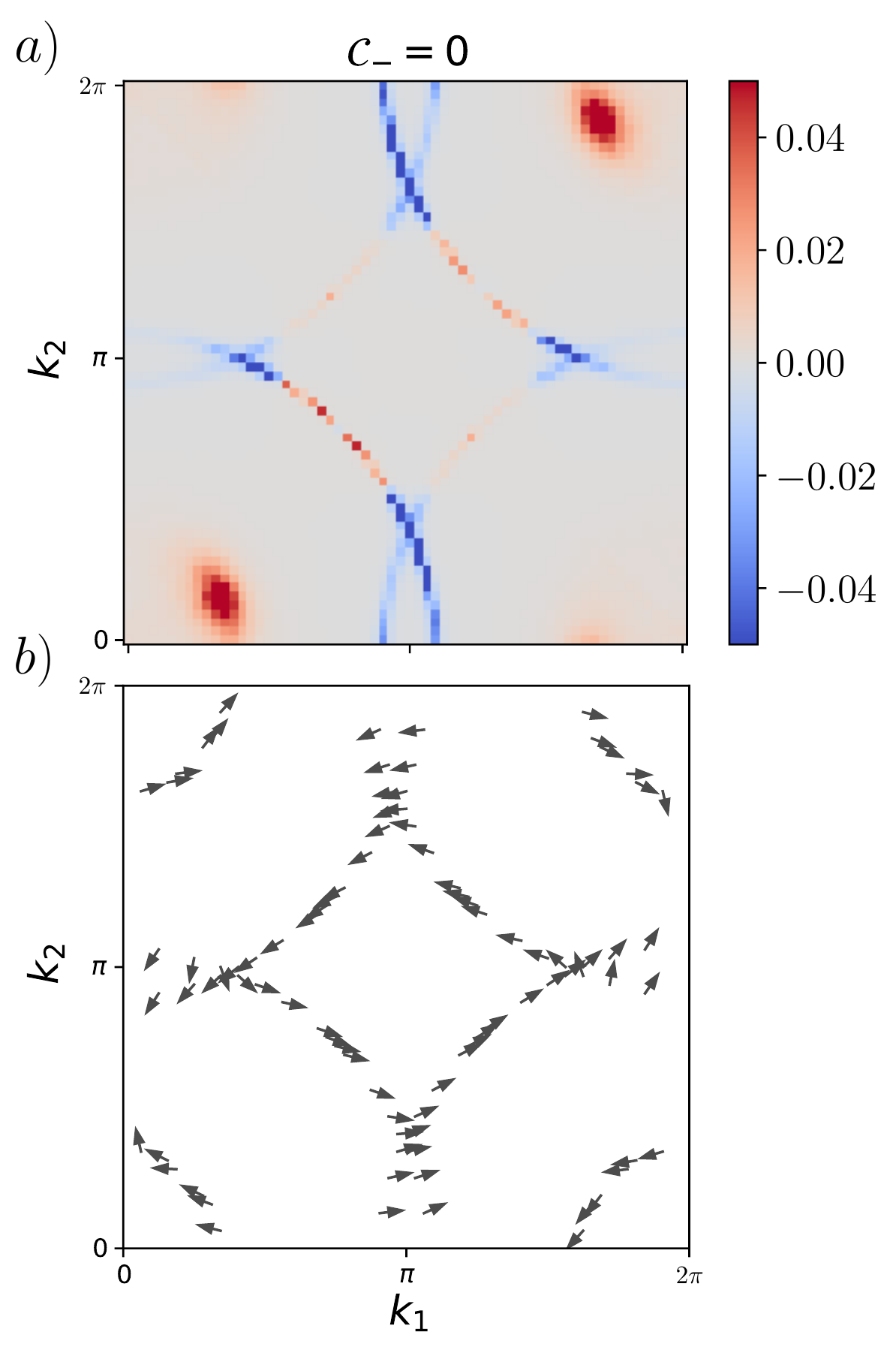} 
\caption{\label{fig:Berry}The a) Berry curvature and b) intraband phase winding are calculated for bands in the pseudospin down sector, for the pairing in Fig.~\ref{fig:maxPeakspwave}. For the folded BZ we can consider 4 pockets, centered around $(k_{1}, k_{2})=(0,0), (0, \pi), (\pi,0), (\pi,\pi)$. Each pocket has $\mathcal{C}_{n,-}=\pm 1$, such that the total Chern number cancels out to $\mathcal{C}_{-}=0$. For each pocket the phase $\phi_{\boldsymbol{k}}$ can be seen to wind in opposite directions for the different pockets.}
\end{figure}

\section{Topological invariant}\label{sec:AppF}
To determine the topological properties of the found odd parity order we consider the pairing in the Bogoliubov-de Gennes (BdG) Hamiltonian
\begin{equation}\label{eq:HBdG}
    H_{\textrm{BdG}} = \sum_{\boldsymbol{k}} \left( c^{\dagger}_{\boldsymbol{k}},  c_{- \boldsymbol{k}} \right) \left( \begin{array}{cc}
        H(\boldsymbol{k}) & \Delta (\boldsymbol{k}) \\
         \Delta^{\dagger} (\boldsymbol{k}) & - H^{\textrm{T}} (-\boldsymbol{k})
    \end{array}\right) \left( \begin{array}{c}  c_{\boldsymbol{k}} \\  c^{\dagger}_{- \boldsymbol{k}}  \end{array} \right)
\end{equation}
with the non-interacting Hamiltonian $H(\boldsymbol{k})$ from Eq.~\eqref{eq:HnonInt}. $\Delta (\boldsymbol{k})$ is the eigenstate of the largest eigenvalue for Eq.~\eqref{eq:SCeq}, scaled to set the minimal gap at the Fermi surface to 0.02eV. As the specific odd parity pairing found in section \ref{sec:OddParity} is block diagonal in pseudospin up and down, the full Hamiltonian can be rearranged into a block-diagonal form
\begin{equation}
    H_{\textrm{BdG}} =  \sum_{\boldsymbol{k}} \boldsymbol{\Psi}_{\boldsymbol{k}}^{\dagger} \left( \begin{array}{cc}
        H_{\textrm{BdG}}^{+}(\boldsymbol{k}) & 0 \\
         0 & H_{\textrm{BdG}}^{-}(\boldsymbol{k})  \end{array}\right) \boldsymbol{\Psi}_{\boldsymbol{k}}
\end{equation}
where $\boldsymbol{\Psi}_{\boldsymbol{k}}= \left( \Psi_{+, \boldsymbol{k}} , \Psi_{-, \boldsymbol{k}} \right)^{\textrm{T}}$ is divided into the pseudospin sectors $ \{(yz, \downarrow), (xz, \downarrow), (xy, \uparrow) \}$ and $\{(yz, \uparrow), (xz, \uparrow), (xy, \downarrow) \}$. The eigenstates and eigenvalues are given as $H_{\textrm{BdG}} |n \rangle = E_{n} |n \rangle$ and for each pseudospin sector $H_{\textrm{BdG}}^{\tau} |n \rangle = E_{n, \tau} |n, \tau \rangle$. We calculate the Berry curvature for all filled bands via~\cite{Berry1984}
\begin{widetext}
\begin{equation}
\begin{aligned}
        \Omega^{(n)}_{z}(\boldsymbol{k})  =& - \textrm{Im} \sum_{n' \neq n }  f(E_{n}) f(1- E_{n'}) \\ & \times \frac{\langle n |\partial_{x} H_{\textrm{BdG}}| n' \rangle  \langle n' | \partial_{y} H_{\textrm{BdG}} | n \rangle - \langle n |\partial_{y} H_{\textrm{BdG}}| n' \rangle  \langle n' | \partial_{x} H_{\textrm{BdG}} | n \rangle}{(E_{n} - E_{n'})^{2}} 
\end{aligned}
\end{equation}
\begin{equation}
\begin{aligned}
   \Omega^{(n)}_{z}(\boldsymbol{k})  = &- \textrm{Im} \sum_{n' \neq n}\sum_{ a, b, c, d} \sum_{i,j \in \{ x,y\} } f(E_{n}) f(1- E_{n'}) \left[ U_{\boldsymbol{k}}^{\dagger} \right]_{n a}  \left[ U_{\boldsymbol{k}}\right]_{b n'}  \left[ U_{\boldsymbol{k}}^{\dagger} \right]_{n' c}   \left[ U_{\boldsymbol{k}} \right]_{d n} \\ & \times \epsilon_{ij} \frac{\langle a |\partial_{i} H_{\textrm{BdG}}| b \rangle  \langle c | \partial_{j} H_{\textrm{BdG}} | d \rangle }{(E_{n} - E_{n'})^{2}} 
\end{aligned}
\end{equation}
\end{widetext}
where $ \epsilon_{xy} = -\epsilon_{yx}=1$ and $\epsilon_{ii}=0$. $| a \rangle$ is the spin-orbital basis $a=(\alpha, \sigma, s)$ with orbital $\alpha$, spin $\sigma$, and sublattice $s$. However, since all $\langle a, + |\partial_{i} H_{\textrm{BdG}}| b, - \rangle =0$ the Berry curvature can be calculated separately for each pseudospin sector
\begin{equation}
    \Omega^{(n)}_{z}(\boldsymbol{k})  = \Omega^{(n, +)}_{z}(\boldsymbol{k})  + \Omega^{(n, -)}_{z}(\boldsymbol{k}) 
\end{equation}
and the Chern number per pseudospin is 
\begin{equation}
\mathcal{C}_{ \tau} = \frac{1}{2 \pi}\int_{\textrm{BZ}} \sum_{n} \Omega^{(n,\tau)}(\boldsymbol{k})  d k_{x} d k_{y}
\end{equation}
In Fig.~\ref{fig:Berry} the Berry curvature for all filled bands is shown for the pseudospin down sector. In addition, the phase $\phi_{\boldsymbol{k}}$ of the intraband pairing, where $\Delta_{\textrm{FS}_{n}}^{\downarrow \downarrow} (\boldsymbol{k}) = |\Delta_{\textrm{FS}_{n}}^{\downarrow \downarrow} (\boldsymbol{k})| e^{i \phi_{\boldsymbol{k}}}$, is plotted along the Fermi surface. The winding of the phase can be seen to be opposite for the bands. If the pockets are fully separated the Chern number for each pocket can also be given by the winding of the phase~\cite{Cheng2010}
\begin{equation}\label{eq:phaseWinding}
    \mathcal{C}_{n} = \frac{1}{2 \pi} \oint _{\textrm{FS}_{n}} \nabla \phi_{\boldsymbol{k}} \cdot d \boldsymbol{k}.
\end{equation}

\bibliography{strainSCrefs}
\end{document}